\documentclass[12pt]{elsart}
\usepackage{epsfig}
\usepackage{pstricks}
\newcommand{\nn}{\noindent}

\def\bo{{\raise.15ex\hbox{\large$\Box$}}}               % D'Alembertian
\def\face{{\raise.2ex\hbox{$\displaystyle \bigodot$}\mskip-2.2mu \llap {$\ddot
        \smile$}}}                                      % happy face
\def\leftrightarrowfill{$\mathsurround=0pt \mathord\leftarrow \mkern-6mu
        \cleaders\hbox{$\mkern-2mu \mathord- \mkern-2mu$}\hfill
        \mkern-6mu \mathord\rightarrow$}       % <--> double differential
\def\dvec#1{\vbox{\ialign{##\crcr
        \leftrightarrowfill\crcr\noalign{\kern-1pt\nointerlineskip}
        $\hfil\displaystyle{#1}\hfil$\crcr}}}           % <--> accent
\def\beq{\begin{equation}}
\def\eeq{\end{equation}}

\def\beqx{\begin{displaymath}}
\def\eeqx{\end{displaymath}}

\def\beqa{\begin{eqnarray}}
\def\eeqa{\end{eqnarray}}

\def \npb#1#2#3{{\rm Nucl.~Phys.} {\bf B#1} (19#2) #3}

\def \stone{{\it B Decays}, edited by S. Stone (World Scientific,
Singapore,1994)}
\newread\epsffilein % file to \read
\newif\ifepsffileok % continue looking for the bounding box?
\newif\ifepsfbbfound % success?
\newif\ifepsfverbose % report what you're making?
\newdimen\epsfxsize % horizontal size after scaling
\newdimen\epsfysize % vertical size after scaling
\newdimen\epsftsize % horizontal size before scaling 
\newdimen\epsfrsize % vertical size before scaling
\newdimen\epsftmp % register for arithmetic manipulation
\newdimen\pspoints % conversion factor
\def\epsfbox#1{\global\def\epsfllx{72}\global\def\epsflly{72}%  
 \global\def\epsfurx{540}\global\def\epsfury{720}%
 \def\lbracket{[}\def\testit{#1}\ifx\testit\lbracket
 \let\next=\epsfgetlitbb\else\let\next=\epsfnormal\fi\next{#1}}%
\def\epsfgetlitbb#1#2 #3 #4 #5]#6{\epsfgrab #2 #3 #4 #5 .\\%
 \epsfsetgraph{#6}}%
\def\epsfnormal#1{\epsfgetbb{#1}\epsfsetgraph{#1}}%
\def\epsfgetbb#1{%
\openin\epsffilein=#1
\ifeof\epsffilein\errmessage{I couldn't open #1, will ignore it}\else
 {\epsffileoktrue \chardef\other=12
 \def\do##1{\catcode`##1=\other}\dospecials \catcode`\ =10
 \loop
 \read\epsffilein to \epsffileline
 \ifeof\epsffilein\epsffileokfalse\else
 \expandafter\epsfaux\epsffileline:. \\%
 \fi
 \ifepsffileok\repeat
 \ifepsfbbfound\else
 \ifepsfverbose\message{No bounding box comment in #1; using
defaults}\fi\fi
 }\closein\epsffilein\fi}%
\def\epsfclipstring{}% do we clip or not? If so,
\def\epsfsetgraph#1{%
 \epsfrsize=\epsfury\pspoints
 \advance\epsfrsize by-\epsflly\pspoints
 \epsftsize=\epsfurx\pspoints
 \advance\epsftsize by-\epsfllx\pspoints
 \epsfxsize\epsfsize\epsftsize\epsfrsize
 \ifnum\epsfxsize=0 \ifnum\epsfysize=0
 \epsfxsize=\epsftsize \epsfysize=\epsfrsize
 \epsfrsize=0pt
 \else\epsftmp=\epsftsize \divide\epsftmp\epsfrsize
 \epsfxsize=\epsfysize \multiply\epsfxsize\epsftmp
 \multiply\epsftmp\epsfrsize \advance\epsftsize-\epsftmp
 \epsftmp=\epsfysize
 \loop \advance\epsftsize\epsftsize \divide\epsftmp 2
 \ifnum\epsftmp>0
 \ifnum\epsftsize<\epsfrsize\else
 \advance\epsftsize-\epsfrsize \advance\epsfxsize\epsftmp \fi
 \repeat
 \epsfrsize=0pt
 \fi
\else \ifnum\epsfysize=0
 \epsftmp=\epsfrsize \divide\epsftmp\epsftsize
 \epsfysize=\epsfxsize \multiply\epsfysize\epsftmp
 \multiply\epsftmp\epsftsize \advance\epsfrsize-\epsftmp
 \epsftmp=\epsfxsize 
 \loop \advance\epsfrsize\epsfrsize \divide\epsftmp 2
 \ifnum\epsftmp>0
 \ifnum\epsfrsize<\epsftsize\else
 \advance\epsfrsize-\epsftsize \advance\epsfysize\epsftmp \fi
 \repeat
 \epsfrsize=0pt
 \else
 \epsfrsize=\epsfysize
 \fi
 \fi
 \ifepsfverbose\message{#1: width=\the\epsfxsize,
height=\the\epsfysize}\fi
 \epsftmp=10\epsfxsize \divide\epsftmp\pspoints
 \vbox to\epsfysize{\vfil\hbox to\epsfxsize{%
 \ifnum\epsfrsize=0\relax
 \includegraphics{#1}%
 \else
 \epsfrsize=10\epsfysize \divide\epsfrsize\pspoints  
 \includegraphics{#1}%
 \fi
 \hfil}}%
\global\epsfxsize=0pt\global\epsfysize=0pt}%
 {\catcode`\%=12
\global\let\epsfpercent=%\global\def\epsfbblit{%BoundingBox}}%
\long\def\epsfaux#1#2:#3\\{\ifx#1\epsfpercent
 \def\testit{#2}\ifx\testit\epsfbblit
 \epsfgrab #3 . . . \\%
 \epsffileokfalse
 \global\epsfbbfoundtrue
 \fi\else\ifx#1\par\else\epsffileokfalse\fi\fi}%
\def\epsfempty{}%
\def\epsfgrab #1 #2 #3 #4 #5\\{%
\global\def\epsfllx{#1}\ifx\epsfllx\epsfempty
 \epsfgrab #2 #3 #4 #5 .\\\else
 \global\def\epsflly{#2}%
 \global\def\epsfurx{#3}\global\def\epsfury{#4}\fi}%
\def\epsfsize#1#2{\epsfxsize}
\def\sss{\scriptscriptstyle}
\def\barp{{\raise.35ex\hbox{${\sss (}$}}---{\raise.35ex\hbox{${\sss )}$}}}
\def\bdbarp{\hbox{$B_d$\kern-1.4em\raise1.4ex\hbox{\barp}}}
\def\bsbarp{\hbox{$B_s$\kern-1.4em\raise1.4ex\hbox{\barp}}}
\def\dbarp{\hbox{$D$\kern-1.1em\raise1.4ex\hbox{\barp}}}

\newcommand{\absvcb}{\vert V_{cb}\vert}

\newcommand{\abseps}{\vert\epsilon\vert}

\newcommand{\fbb}{f^2_{B_d}\hat{B}_{B_d}}
\newcommand{\fbbs}{f^2_{B_s}\hat{B}_{B_s}}
\newcommand{\fbd}{f_{B_d}}
\newcommand{\fbs}{f_{B_s}}

\def\rly#1{\mathrel{\raise.3ex\hbox{$#1$\kern-.75em\lower1ex\hbox{$\sim$}}}}

\def\mt{m_t}

\def\mc{m_c}

\newcommand{\delmd}{\Delta M_d} 
\newcommand{\delms}{\Delta M_s}

\newcommand{\kkbar}{$K^0$--${\overline{K^0}}$}
\newcommand{\bdbdbar}{$B_d^0$--${\overline{B_d^0}}$}
\newcommand{\bsbsbar}{$B_s^0$--${\overline{B_s^0}}$}
\begin{document}

\begin{frontmatter}
\title{
{\normalsize
\textnormal{DESY 99-083} \hspace{\fill} \mbox{ }\\
\textnormal{UdeM-GPP-TH-99-60} \hspace{\fill} \mbox{ }\\
\textnormal{May 1999} \hspace{\fill} \mbox{ }\\[5ex]
}
PRECISION FLAVOUR PHYSICS AND SUPERSYMMETRY\thanksref{okun}}
\thanks[okun]{Contribution to the Festschrift for L.~B.~Okun, 
to appear in a special issue of Physics Reports, 
eds.~V.~L.~Telegdi and K.~Winter}
\author{Ahmed Ali}
\address{Deutsches Elektronen-Synchrotron DESY, Hamburg, Germany}

\author{David London}
\address{Laboratoire Ren\'e J.-A. L\'evesque, Universit\'e de
Montr\'eal, \\ C.P. 6128, succ.\ centre-ville,
 Montr\'eal, QC, Canada H3C 3J7}

\begin{abstract}
\noindent
We review the salient features of a comparative study of the profile
of the CKM unitarity triangle, and the resulting CP-violating phases
$\alpha$, $\beta$ and $\gamma$ in $B$ decays, in the standard model
and in several variants of the minimal supersymmetric standard model
(MSSM), reported recently by us. These theories are characterized by a
single phase in the quark flavour mixing matrix and give rise to
well-defined contributions in the flavour-changing-neutral-current
transitions in $K$ and $B$ decays.  We analyse the supersymmetric
contributions to the mass differences in the \bdbdbar\ and \bsbsbar\ 
systems, $\Delta M_d$ and $\Delta M_s$, respectively, and to the
CP-violating quantity $\abseps$ in $K$ decays. Our analysis shows that
the predicted ranges of $\beta$ in the standard model and in MSSM
models are very similar. However, precise measurements at
$B$-factories and hadron machines may be able to distinguish these
theories in terms of the other two CP-violating phases $\alpha$ and
$\gamma$.
\end{abstract}

\end{frontmatter}

\newpage

\section{Introduction}
In this article, written to honour the scientific achievements of Lev
Okun, we discuss some selected topics in quark flavour physics. In
particular, we review the present status of quark flavour mixing in
the Standard Model (SM) and in some variants of the Minimal
Supersymmetric Standard Model (MSSM). The idea is to present
contrasting profiles of quark flavour physics in these theoretical
frameworks which can be tested in the next generation of experiments
in flavour physics. The emphasis in this paper is on CP violating
asymmetries and particle-antiparticle mixings induced by weak
interactions. These topics are close to Lev Okun's own scientific
research. In fact, the possibility of observing violation of
CP-invariance in heavy particle decays was proposed in an important
paper by Okun, Pontecorvo and Zakharov in 1975, just after the
discovery of the charmed hadrons~\cite{OPZ-75}. To be specific, these
authors studied the consequences of $D^0\overline{D^0}$ pair
production, subsequent $D^0$-$\overline{D^0}$ mixing, and CP violation
for the final states involving same-sign $\ell^\pm \ell^\pm$ and
opposite-sign $\ell^+ \ell^-$ dileptons. In particular, as a measure
of $D^0$-$\overline{D^0}$ mixing they proposed the measurement of the
ratio of the same-sign to the inclusive dilepton events,
\beq 
R_D \equiv \frac{N^{++} + N^{--}}{N^{+-} + N^{-+} + N^{++} + N^{--}} =
\frac{1}{2} \frac{(\Gamma_S - \Gamma_L)^2 + 4(\Delta M_{SL})^2}{
  (\Gamma_S + \Gamma_L)^2 + 4(\Delta M_{SL})^2} ~,
\eeq 
where $\Gamma_S$ and $\Gamma_L$ are the widths of the (short-lived)
$D_S$ and (long-lived) $D_L$ mesons, respectively, and $\Delta M_{SL}$
is their mass difference. They also suggested the measurement of the
charge asymmetry
\beq 
\delta_D \equiv \frac{N^{++}-N^{--}}{N^{++}+N^{--}} \simeq 4 \mbox{Re}
~\epsilon_D ~,
\eeq 
as a measure of CP violation. Here, $\epsilon_D$ is the CP-violating
parameter in the wave-functions of $D_S$ and $D_L$ mesons, analogous to
the corresponding parameter $\epsilon_K$ in the $K$-system
\cite{PDG98}, 
\beq 
D_S \sim D_1 + \epsilon_D D_2, ~~~~~D_L \sim D_2 + \epsilon_D D_1~,
\eeq 
where $D_1$ and $D_2$ are the pure CP states.

So far, neither $R_D$ nor $\delta_D$ have been measured \cite{PDG98}.
In fact, in the Cabibbo-Kobayashi-Maskawa (CKM) theory of quark
flavour mixing \cite{CKM}, which is now an integral part of the SM, no
measurable effects are foreseen for either of the ratios $R_D$ and
$\delta_D$, due to the experimentally established hierarchies in the
quark mass spectrum and the CKM matrix elements. Typically, one has in
the SM \cite{Donoghue86},
\beq
\frac{\Delta M_{SL}}{\Gamma_S + \Gamma_L} \simeq O(10^{-5}), 
~~~\frac{\Gamma_S - \Gamma_L}{\Gamma_S + \Gamma_L} \ll 1 ~,
\eeq
with $\delta_D$ completely negligible.  By virtue of this, the
quantities $R_D$ and $\delta_D$ have come to be recognized as useful
tools to search for physics beyond the SM \cite{BSN95,Hewett96}.

The OPZ formulae also apply to the time-integrated effects of mixing
and CP violation in the \bdbdbar\ and \bsbsbar\ systems, and they were
used in the analysis \cite{Eurojet} of the UA1 data on inclusive
dilepton production \cite{UA1BB}. Calling the corresponding mixing
measures $R_{B_d}$ and $R_{B_s}$, respectively, present experiments
yield $R_{B_d} \simeq 0.17$ and $R_{B_s} \simeq 1/2$ \cite{PDG98}.
These measurements are consistent with the more precise time-dependent
measurements, yielding $\Delta M_{B_d}=0.471\pm 0.016$ (ps)$^{-1}$
\cite{Alexander98} and the 95\% C.L. upper limit $\Delta M_{B_s} >
12.4$ (ps)$^{-1}$ \cite{Parodiconf98}.  However, the corresponding
CP-violating charge asymmetries $\delta_{B_d}$ and $\delta_{B_s}$ in
the two neutral $B$-meson systems have not been measured, with the
present best experimental limit being $\delta_{B_d} =0.002\pm 0.007
\pm 0.003$ from the OPAL collaboration \cite{PDG98} and no useful
limit for the quantity $\delta_{B_s}$. These charge asymmetries are
expected to be very small in the SM, reflecting essentially that the
width and mass differences $\Delta \Gamma$ and $\Delta M$ in the
\bdbdbar\ and \bsbsbar\ complexes are relatively real.  Typical
estimates in the SM are in the range $\delta_{B_d}=O(10^{-3})$ and
$\delta_{B_s}=O(10^{-4})$.  Hence, like $\delta_D$, they are of
interest in the context of physics beyond the SM
\cite{Aliaydin79,Randallsu}.

With the advent of $B$ factories and HERA-B, one expects that a large
number of CP asymmetries in partial decay rates of $B$ hadrons and
rare $B$ decays will become accessible to experimental and theoretical
studies. Of particular interest in this context are the
flavour-changing neutral-current (FCNC) processes which at the quark
level can be thought of as taking place through induced $b \to d$ and
$b \to s$ transitions.  In terms of actual laboratory measurements,
these FCNC processes will lead to $\Delta B=1$, $\Delta Q=0$ decays
such as $B \to (X_s, X_d) l^+ l^-$ and $B \to (X_s, X_d) \gamma$,
where $X_s(X_d)$ represents an inclusive hadronic state with an
overall quantum number $S=\pm 1(0)$, as well as their exclusive decay
counterparts, such as $B \to (K,K^*,\pi,\rho,...)  \ell^+ \ell^-$ and
$B \to (K^*,\rho,\omega,...) \gamma$. Of these, the decays $B \to X_s
\gamma$ and $B \to K^* \gamma$ have already been measured
\cite{PDG98}.  The $\Delta B=2$, $\Delta Q=0$ transitions lead to
\bdbdbar\ and \bsbsbar\ mixings, briefly discussed above.  Likewise,
non-trivial bounds have been put on the CP-violating phase $\sin 2
\beta$ from the time-dependent rate asymmetry in the decays
$B^0/\overline{B^0} \to J/\psi K_s$ \cite{CDF99}. In $K$ decays, the
long sought after effect involving direct CP violation has been
finally established through the measurement of the ratio
$\epsilon^\prime/\epsilon$ \cite{KTEV99,NA31}. This and the
measurement of the CP-violating quantity $\abseps$ in $K_L \to \pi
\pi$ decays \cite{PDG98} represent the $s \to d$ FCNC transitions.
Likewise, there exists great interest in the studies of FCNC rare $K$
decays such as $K^+ \to \pi^+ \nu \bar{\nu}$ and $K_L \to \pi^0 \nu
\bar{\nu}$ \cite{Burasreview98}, of which a single event has been
measured in the former decay mode \cite{BNL787}.

The FCNC processes and CP asymmetries in $K$ and $B$ decays provide
stringent tests of the SM. The short-distance contributions to these
transitions are dominated by the top quark, and hence these decays and
asymmetries provide information on the weak mixing angles and phases
in the matrix elements $V_{td}$, $V_{ts}$ and $V_{tb}$ of the CKM
matrix. Some information on the last of these matrix elements is also
available from the direct production and decay of the top quarks at
the Fermilab Tevatron \cite{CDFvtb}. The measurement of $V_{tb}$ will
become quite precise at the LHC and linear colliders. Moreover, with
advances in determining the (quark) flavour of a hadronic jet, one
also anticipates being able to measure the matrix element $V_{ts}$
(and possibly also $V_{td}$).
 
We shall concentrate here on the analysis of the data at hand and in
forthcoming experiments which will enable us to test precisely the
unitarity of the CKM matrix.  These tests will be carried out in the
context of the Unitarity Triangles (UT). The sides of UTs will be
measured in $K$ and $B$ decays and the angles of these UTs will be
measured by CP asymmetries. Consistency of a theory, such as the SM,
requires that the two sets of independent measurements yield the same
values of the CKM parameters, or, equivalently the CP-violating phases
$\alpha$, $\beta$ and $\gamma$. We are tacitly assuming that there is
only one CP-violating phase in weak interactions. This is the case in
the SM but also in a number of variants of Supersymmetric Models,
which, however, do have additional contributions to the FCNC
amplitudes.  In fact, it is the possible effect of these additional
contributions which will be tested.  In this case, quantitative
predictions can be made which, in principle, allow experiments to
discriminate among these theories \cite{AL-99}. As we shall see, the
case for distinguishing the SM and the MSSM rests on the experimental
and theoretical precision that can achieved in various input
quantities. Of course, there are many other theoretical scenarios in
which deviations from the pattern of flavour violation in the SM are
not minimal. For example, in the context of supersymmetric models, one
may have non-diagonal quark-squark-gluino couplings, which also
contain additional phases. These can contribute significantly to the
magnitude and phase of $b \to d$, $b \to s$ and $s \to d$ transitions,
which would then violate the SM flavour-violation pattern rather
drastically. In this case it is easier to proclaim large deviations
from the SM but harder to make quantitative predictions.

This paper is organized as follows. In Section 2, we discuss the
profile of the unitarity triangle within the SM. We describe the input
data used in the fits and present the allowed region in $\rho$--$\eta$
space, as well as the presently-allowed ranges for the CP angles
$\alpha$, $\beta$ and $\gamma$. We also discuss the fits in the superweak 
scenario, which differs from the SM fits in that we no longer use the
constraint from the CP-violating quantity $\abseps$. The superweak fits
are not favoured by the data and we quantify this in terms of the 95\% C.L.
exclusion contours.
 We turn to supersymmetric models in
Section 3. We review several variants of the MSSM, in which the new
phases are essentially zero. Restricting ourselves to flavour
violation in charged-current transitions, we include the effects of
charged Higgses $H^\pm$, a light scalar top quark (assumed here
right-handed as suggested by the precision electroweak fits) and
chargino $\chi^\pm$. In this scenario, which covers an important part
of the SUSY parameter space, the SUSY contributions to \kkbar,
\bdbdbar\ and \bsbsbar\ mixing are of the same form and can be
characterized by a single parameter $f$.  Including the NLO
corrections in such models, we compare the profile of the unitarity
triangle in SUSY models, for various values of $f$, with that of the
SM.  We conclude in Section 4.

\section{Unitarity Triangle: SM Profile}

\begin{figure}
\vskip -1.0truein
\centerline{\epsfxsize 3.5 truein \epsfbox {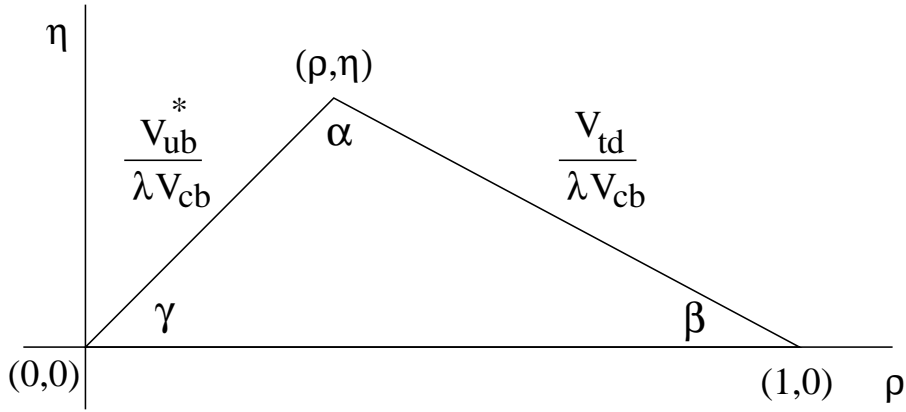}}
\vskip -1.2truein
\caption{\it The unitarity triangle. The angles $\alpha$, $\beta$ and 
$\gamma$ can be measured via CP violation in the $B$ system.}
\label{triangle}
\end{figure}

Within the standard model (SM), CP violation is due to the presence of
a nonzero complex phase in the Cabibbo-Kobayashi-Maskawa (CKM) quark  
mixing matrix $V$ \cite{CKM}. A particularly useful parametrization of
the CKM matrix, due to Wolfenstein \cite{Wolfenstein}, follows from
the observation that the elements of this matrix exhibit a hierarchy  
in terms of $\lambda$, the Cabibbo angle. In this parametrization the
CKM matrix can be written approximately as
\beq
V \simeq \left(\matrix{
 1-{1\over 2}\lambda^2 & \lambda
 & A\lambda^3 \left( \rho - i\eta \right) \cr
 -\lambda ( 1 + i A^2 \lambda^4 \eta )
& 1-{1\over 2}\lambda^2 & A\lambda^2 \cr
 A\lambda^3\left(1 - \rho - i \eta\right) & -A\lambda^2 & 1 \cr}\right)~.
\label{CKM}
\eeq
The allowed region in $\rho$--$\eta$ space can be elegantly displayed  
using the so-called unitarity triangle (UT). The unitarity of the CKM
matrix leads to the following relation:
\beq
V_{ud} V_{ub}^* + V_{cd} V_{cb}^* + V_{td} V_{tb}^* = 0~.
\eeq
Using the form of the CKM matrix in Eq.~(\ref{CKM}), this can be recast as
\beq
\label{trianglerel}
\frac{V_{ub}^*}{\lambda V_{cb}} + \frac{V_{td}}{\lambda V_{cb}} = 1~,
\eeq
which is a triangle relation in the complex plane (i.e.\ $\rho$--$\eta$
space), illustrated in Fig.~\ref{triangle}. Thus, allowed values of $\rho$
and $\eta$ translate into allowed shapes of the unitarity triangle.

% This is Figure 1

Constraints on $\rho$ and $\eta$ come from a variety of sources. Of
the quantities shown in Fig.~\ref{triangle}, $|V_{cb}|$ and $|V_{ub}|$
can be extracted from semileptonic $B$ decays, while $|V_{td}|$ is
probed in \bdbdbar\ mixing. The interior CP-violating angles $\alpha$,
$\beta$ and $\gamma$ can be measured through CP asymmetries in $B$
decays \cite{BCPasym}. Additional constraints come from CP violation
in the kaon system ($\abseps$), as well as \bsbsbar\ mixing. As
the constraints that are expected to come from the
rare $B$ and $K$ decays mentioned earlier are not of interest for the CKM 
phenomenology at present, we shall not include them in our fits.

\subsection{Input Data}

The CKM matrix as parametrized in Eq.~(\ref{CKM}) depends on four
parameters: $\lambda$, $A$, $\rho$ and $\eta$. We summarize below the
experimental and theoretical data which constrain these CKM
parameters.
\begin{itemize}
\item $\vert V_{us}\vert$, $\vert V_{cb}\vert$ and $\vert 
V_{ub}/V_{cb}\vert$:
\end{itemize}
We recall that $\vert V_{us}\vert$ has been extracted with good
accuracy from $K\to\pi e\nu$ and hyperon decays \cite{PDG98} to be
$\vert V_{us}\vert=\lambda=0.2196\pm 0.0023$.  The determination of
$\absvcb$ is based on the combined analysis of the inclusive and
exclusive $B$ decays: $ \vert V_{cb} \vert = 0.0395 \pm 0.0017$
\cite{PDG98}, yielding $A = 0.819 \pm 0.035$. The knowledge of the CKM
matrix element ratio $|V_{ub}/V_{cb}|$ is based on the analysis of the
end-point lepton energy spectrum in semileptonic decays $B \to X_{u}
\ell \nu_\ell$ and the measurement of the exclusive semileptonic
decays $B \to (\pi, \rho) \ell \nu_\ell$. Present measurements in both
the inclusive and exclusive modes are compatible with $\left\vert
  V_{ub}/V_{cb} \right\vert = 0.093\pm 0.014$ \cite{Parodiconf98}.
This gives $\sqrt{\rho^2 + \eta^2} = 0.423 \pm 0.064$.
\begin{itemize}
\item {$ \abseps, \hat{B}_K$}:
\end{itemize}
 The experimental value of
$\abseps$ is \cite{PDG98}:
\beq
\abseps = (2.280\pm 0.013)\times 10^{-3}~.
\eeq
In the standard model, $\abseps$ is essentially proportional to the
imaginary part of the box diagram for \kkbar\ mixing and is given by
\cite{Burasetal}
\begin{eqnarray}
\abseps &=& \frac{G_F^2f_K^2M_KM_W^2}{6\sqrt{2}\pi^2\Delta M_K}
\hat{B}_K\left(A^2\lambda^6\eta\right)
\bigl(y_c\left\{\hat{\eta}_{ct}f_3(y_c,y_t)-\hat{\eta}_{cc}\right\}
 \nonumber \\
&~& ~~~~~~~~~~~~~~+ 
~\hat{\eta}_{tt}y_tf_2(y_t)A^2\lambda^4(1-\rho)\bigr), 
\label{eps}
\end{eqnarray}
where $y_i\equiv m_i^2/M_W^2$, and the functions $f_2$ and $f_3$
are the Inami-Lim functions \cite{InamiLim}.
Here, the $\hat{\eta}_i$ are QCD correction factors, calculated at
next-to-leading order: ($\hat{\eta}_{cc}$) \cite{HN94},
($\hat{\eta}_{tt}$) \cite{etaB} and ($\hat{\eta}_{ct}$) \cite{HN95}.
The theoretical uncertainty in the expression for $\abseps$ is in the
renormalization-scale independent parameter $\hat{B}_K$, which
represents our ignorance of the matrix element $\langle K^0
\vert {({\overline{d}}\gamma^\mu (1-\gamma_5)s)}^2 \vert
{\overline{K^0}}\rangle$. Recent calculations of $\hat{B}_K$ using
lattice QCD methods are summarized at the 1998 summer conferences by Draper 
\cite{Draper98} and Sharpe \cite{Sharpe98}, yielding
\begin{equation}
 \hat{B}_K=0.94 \pm 0.15 .
\label{BKrange}
\end{equation}
\begin{itemize}
\item {$ \Delta M_d, \fbb$}:
\end{itemize}
 The present world average for
  $\Delta M_d$ is \cite{Alexander98}
\beq
\Delta M_d = 0.471 \pm 0.016~(ps)^{-1} ~.
\eeq
The mass difference $\Delta M_d$ is calculated from the \bdbdbar\ box
diagram, dominated by $t$-quark exchange:
\beq
\label{bdmixing}
\Delta M_d = \frac{G_F^2}{6\pi^2}M_W^2M_B\left(\fbb\right)\hat{\eta}_B y_t
f_2(y_t) \vert V_{td}^*V_{tb}\vert^2~, \label{xd}
\eeq
where, using Eq.~(\ref{CKM}), $\vert V_{td}^*V_{tb}\vert^2=
A^2\lambda^{6} [\left(1-\rho\right)^2+\eta^2]$. Here, $\hat{\eta}_B$
is the QCD correction, which has 
the value $\hat{\eta}_B=0.55$, calculated in the $\overline{MS}$
scheme \cite{etaB}.

For the $B$ system, the hadronic uncertainty is given by $\fbb$.
 Present estimates of this quantity using lattice QCD yield $\fbd
\sqrt{\hat{B}_{B_d}} =(190 \pm 23)$ MeV in the quenched approximation
\cite{Draper98,Sharpe98}.
The effect of unquenching is not yet understood completely. Taking the
MILC collaboration estimates of unquenching would increase the central
value of $\fbd \sqrt{\hat{B}_{B_d}}$ by $21$ MeV \cite{MILC98}. In the
fits discussed here \cite{AL-99}, the following range has been used:
\begin{equation}
\fbd \sqrt{\hat{B}_{B_d}} = 215 \pm 40 ~\mbox{MeV}~.
\label{FBrange}
\end{equation}
\begin{itemize}
\item {$ \Delta M_s, \fbbs$}:
\end{itemize}
 The \bsbsbar\ box
  diagram is again dominated by $t$-quark exchange, and the mass
  difference between the mass eigenstates $\delms$ is given by a
  formula analogous to that of Eq.~(\ref{xd}):
\beq
\delms = \frac{G_F^2}{6\pi^2}M_W^2M_{B_s}\left(\fbbs\right)
\hat{\eta}_{B_s} y_t f_2(y_t) \vert V_{ts}^*V_{tb}\vert^2~.
\label{xs}
\eeq
Using the fact that $\vert V_{cb}\vert=\vert V_{ts}\vert$ (Eq.~(\ref{CKM})),
it is clear that one of the sides of the unitarity triangle, $\vert
V_{td}/\lambda V_{cb}\vert$, can be obtained from the ratio of $\delmd$ and
$\delms$,
\beq
\frac{\delms}{\delmd} =
 \frac{\hat{\eta}_{B_s}M_{B_s}\left(\fbbs\right)}
{\hat{\eta}_{B_d}M_{B_d}\left(\fbb\right)}
\left\vert \frac{V_{ts}}{V_{td}} \right\vert^2.
\label{xratio}
\eeq
The only real uncertainty in this quantity is the ratio of hadronic
matrix elements $\fbbs/\fbb$. The present estimate of this quantity is
\cite{Draper98,Sharpe98}
\beq
\label{xirange}
\xi_s=1.14\pm 0.06 ~.
\eeq
The present lower bound on $\Delta M_s$ is: $\Delta M_s > 12.4
~\mbox{(ps)}^{-1}$ (at $95\%$ C.L.) \cite{Parodiconf98}.

\begin{table}
\hfil
\vbox{\offinterlineskip
\halign{&\vrule#&
 \strut\quad#\hfil\quad\cr
\noalign{\hrule}
height2pt&\omit&&\omit&\cr
& Parameter && Value & \cr
height2pt&\omit&&\omit&\cr
\noalign{\hrule}
height2pt&\omit&&\omit&\cr
\footnotesize
& $\lambda$ && $0.2196$ & \cr
& $\vert V_{cb} \vert $ && $0.0395 \pm 0.0017$ & \cr
& $\vert V_{ub} / V_{cb} \vert$ && $0.093 \pm 0.014$ & \cr
& $\abseps$ && $(2.280 \pm 0.013) \times 10^{-3}$ & \cr
& $\Delta M_d$ && $(0.471 \pm 0.016)~(ps)^{-1}$ & \cr
& $\Delta M_s$ && $ > 12.4 ~(ps)^{-1}$ & \cr 
& $\overline{\mt}(\mt(pole))$ && $(165 \pm 5)$ GeV & \cr
& $\overline{\mc}(\mc(pole))$ && $1.25 \pm 0.05$ GeV & \cr
& $\hat{\eta}_B$ && $0.55$ & \cr
& $\hat{\eta}_{cc} $ && $1.38 \pm 0.53$ & \cr
& $\hat{\eta}_{ct} $ && $0.47 \pm 0.04$ & \cr
& $\hat{\eta}_{tt} $ && $0.57$ & \cr
& $\hat{B}_K$ && $0.94 \pm 0.15$ & \cr
& $\fbd\sqrt{\hat{B}_{B_d}} $ && $215 \pm 40$ MeV & \cr
& $\xi_s $ && $1.14 \pm 0.06$  & \cr
height2pt&\omit&&\omit&\cr
\noalign{\hrule}}}
\caption{\it Data and theoretical input used in the CKM fits.}
\label{datatable}
\end{table}

There are two other measurements which should be mentioned here.
First, the KTEV collaboration \cite{KTEV99} has recently reported a
measurement of direct CP violation in the $K$ sector through the ratio
$\epsilon^\prime/\epsilon$, with
\beq
{\rm Re} (\epsilon^\prime/\epsilon) = \left( 28.0 \pm 3.0 (\mbox{stat}) 
\pm 2.6 (\mbox{syst}) \pm 1.0(\mbox{MC stat})\right) \times 10^{-4} ~,
\eeq
in agreement with the earlier measurement by the CERN experiment NA31
\cite{NA31}, which reported a value of $(23 \pm 6.5) \times 10^{-4}$
for the same quantity. The present world average is ${\rm Re}
(\epsilon^\prime/\epsilon) =(21.8 \pm 3.0) \times 10^{-4}$. This
combined result excludes the superweak model \cite{superweak} by more
than $7\sigma$. 

A great deal of theoretical effort has gone into calculating this
quantity at next-to-leading order accuracy in the SM
\cite{Buraseps,Martinellieps,Bertolinieps}. The result of this
calculation can be summarized in the following form due to Buras and
Silvestrini \cite{BS98}:
\beq
{\rm Re} (\epsilon^\prime/\epsilon) = {\rm Im} \lambda_t 
\left[ -1.35 + R_s \left( 1.1 \vert r_Z^{(8)} \vert B_6^{(1/2)} 
+ (1.0 -0.67 \vert r_Z^{(8)} \vert) B_8^{(3/2)} \right) \right].
\label{epsilonp}
\eeq
Here $\mbox{Im}~\lambda_t= \mbox{Im}~V_{td}V_{ts}^* = A^2 \lambda^5 \eta$,
and $r_Z^{(8)}$
represents the short-distance contribution, which at the NLO precision
is estimated to lie in the range $ 6.5 \leq \vert r_Z^{(8)} \vert \leq
8.5$ \cite{Buraseps,Martinellieps}. The quantities
$B_6^{(1/2)}=B_6^{(1/2)}(m_c)$ and $B_8^{(3/2)}=B_8^{(3/2)}(m_c)$ are
the matrix elements of the $\Delta I=1/2$ and $\Delta I =3/2$
operators $O_6$ and $O_8$, respectively, calculated at the scale
$\mu=m_c$. Lattice-QCD \cite{B68-Latt} and the $1/N_c$ expansion
\cite{BurasNc} yield:
\beq
0.8 \leq B_6^{(1/2)} \leq 1.3,~~~~~~~~~~~0.6 \leq B_8^{(3/2)} \leq 1.0~.
\eeq
Finally, the quantity $R_s$ in Eq.~(\ref{epsilonp}) is defined as:
\beq
R_s \equiv \left( \frac{150~\mbox{MeV}}{m_s(m_c) + m_d(m_c)} \right)^2 ~,
\label{Rsdef}
\eeq
essentially reflecting the $s$-quark mass dependence. The present
uncertainty on the CKM matrix element is $\pm 23\%$, which is already
substantial. However, the theoretical uncertainties related to the
other quantities discussed above are considerably larger. For
example, the ranges $\epsilon^\prime/\epsilon=(5.3 \pm 3.8) \times
10^{-4}$ and $\epsilon^\prime/\epsilon=(8.5 \pm 5.9) \times 10^{-4}$,
assuming $m_s(m_c)=150\pm 20$ MeV and $m_s(m_c)=125\pm 20$ MeV,
respectively, have been quoted as the best representation of the
status of $\epsilon^\prime/\epsilon$ in the SM \cite{Burasreview98}.
These estimates are somewhat on the lower side compared to the data
but not inconsistent.

Thus, whereas $\epsilon^\prime/\epsilon$ represents a landmark
measurement, establishing for the first time direct CP-violation 
in decay amplitudes, and  hence removing the superweak model of 
Wolfenstein and its  various incarnations from further consideration, its 
impact on CKM phenomenology,
particularly in constraining the CKM parameters, is marginal. For
this reason, the measurement of
$\epsilon^\prime/\epsilon$ is not included in the CKM fits 
summarized here.

Second, the CDF collaboration has recently made a measurement of $\sin
2\beta$ \cite{CDF99}. In the Wolfenstein parametrization, 
$-\beta$ is
the phase of the CKM matrix element $V_{td}$. From Eq.~(\ref{CKM}) one
can readily find that
\beq
\sin (2 \beta) = \frac{2\eta(1-\rho)}{(1-\rho)^2 + \eta^2} ~.
\eeq
Thus, a measurement of $\sin 2\beta$ would put a strong contraint on
the parameters $\rho$ and $\eta$. However, the CDF measurement gives
\cite{CDF99}
\beq
\sin 2\beta = 0.79^{+0.41}_{-0.44} ~,
\eeq
or $\sin 2\beta > 0$ at 93\% C.L. 
This constraint is quite weak -- the indirect measurements already
constrain $0.52 \le \sin 2\beta \le 0.94$ at the 95\% C.L.\ in the SM
\cite{AL-99}.
(The CKM fits reported recently in the 
literature \cite{Parodiconf98,Mele98,PRS98} yield
similar ranges.)  In light of this, this measurement is not included in 
the fits. The data used in the CKM fits are summarized in Table 
\ref{datatable}.
\subsection{SM Fits}

\begin{figure}
\vskip -1.0truein
\centerline{\epsfxsize 3.5 truein \epsfbox {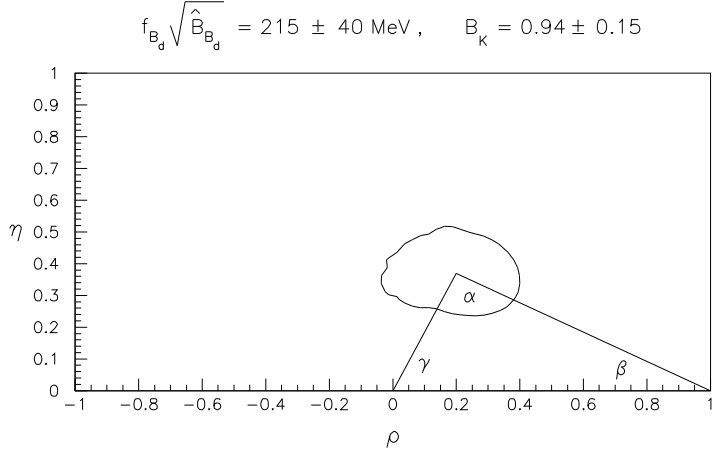}}
\vskip -1.4truein
\caption{\it Allowed region in $\rho$--$\eta$ space in the SM, from a
  fit to the ten parameters discussed in the text and given in Table
  \protect{\ref{datatable}}. The limit on $\Delta M_s$ is included
  using the amplitude method \protect\cite{Moser97}. The
  theoretical errors on $\fbd\protect\sqrt{\hat{B}_{B_d}}$,
  $\hat{B}_K$ and $\xi_s$ are treated as Gaussian. The solid line
  represents the region with $\chi^2=\chi_{min}^2+6$ corresponding to
  the 95\% C.L.\ region. The triangle shows the best fit.
 (From Ref.~\protect\cite{AL-99}.)}
 \label{rhoeta1}
\end{figure}

 In
the fit presented here \cite{AL-99}, ten parameters are allowed to vary: 
$\rho$, $\eta$, $A$, 
$m_t$, $m_c$, $\eta_{cc}$, $\eta_{ct}$, $f_{B_d} \sqrt{\hat{B}_{B_d}}$,
$\hat{B}_K$, and $\xi_s$. The $\Delta M_s$ constraint is included using
the amplitude method \cite{Moser97}. The rest of the parameters are fixed to 
their central values.  
The allowed (95\% C.L.) $\rho$--$\eta$ region is shown in
Fig.~\ref{rhoeta1}.
The best-fit values of the CKM parameters are:
\beq
\lambda = 0.2196 ~~\mbox{(fixed)}, ~~A = 0.817, ~~\rho = 0.196, ~~\eta = 
0.37 ~.
 \eeq
The ``best-fit'' values of the CKM matrix elements are as follows
(Note that we have rounded all elements except $V_{ub}$ and $V_{td}$ to the 
nearest 0.005): 
\beq
V \simeq \left(\matrix{
 0.975 & 0.220
 & 0.002 - 0.003 i \cr
 -0.220
& 0.975 & 0.040 \cr
0.007 - 0.003 i & -0.040 & 1 \cr}\right)~.
\label{CKMfit}
\eeq
Now, turning to the ratios of CKM matrix elements, which one comes across
in the CKM phenomenology, the ``best-fit" values are:
\beq
\left\vert\frac{V_{td}}{V_{ts}}\right\vert = 0.19,
~~\left\vert\frac{V_{td}}{V_{ub}}\right\vert = 2.12,
~~\left\vert\frac{V_{ub}}{V_{cb}}\right\vert= 0.091~.
\label{Ratio-CKM}
\eeq
The $95\%$ C.L.  ranges are:
\beq
0.15 \leq \left\vert\frac{V_{td}}{V_{ts}}\right\vert \leq  0.24 ~,~~ 
1.30 \leq \left\vert\frac{V_{td}}{V_{ub}}\right\vert \leq 3.64 ~,~~
0.06 \leq \left\vert\frac{V_{ub}}{V_{cb}}\right\vert \leq  0.125 ~.
\label{Ratio-CKM-ranges}
\eeq

With the above fits, the ``best-fit" value of $\delms$ is 
$\delms = 16.6$ ~(ps)$^{-1}$. The corresponding
$95\%$ C.L.  allowed range is:
\beq
   12.4~\mbox{(ps)}^{-1} \leq \delms \leq 27.9~\mbox{(ps)}^{-1}.
\eeq

% This is Figure 2

The CP angles $\alpha$, $\beta$ and $\gamma$ can be measured in
CP-violating rate asymmetries in $B$ decays. These angles can be
expressed in terms of $\rho$ and $\eta$. Thus, different shapes of the
unitarity triangle are equivalent to different values of the CP
angles. Referring to Fig.~\ref{rhoeta1}, we note that the preferred
(central) values of these angles are $(\alpha,\beta,\gamma) =
(93^\circ,25^\circ,62^\circ)$. The allowed ranges at 95\% C.L. are
\beq
\label{CPangleregion}
65^\circ \le \alpha \le 123^\circ ~,~~
16^\circ \le \beta \le 35^\circ ~,~~
36^\circ \le \gamma \le 97^\circ ~.
\eeq
These ranges are similar to the ones obtained in \cite{Mele98,PRS98}, but
not identical as the input parameters differ.

\begin{figure}
\vskip -1.0truein
\centerline{\epsfxsize 3.5 truein \epsfbox {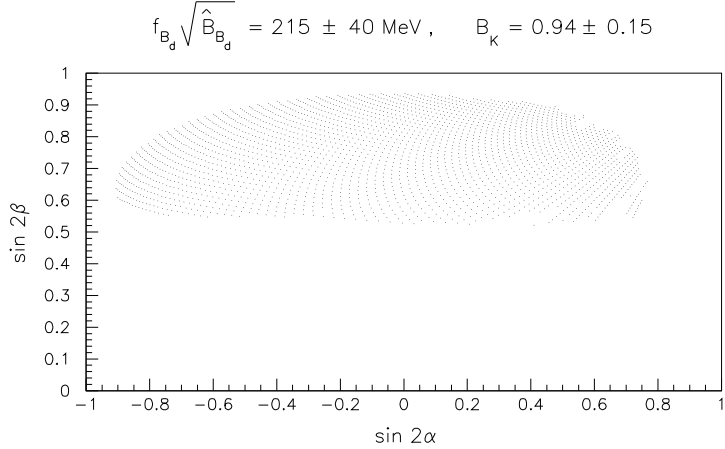}}
\vskip -1.4truein
\caption{\it Allowed 95\% C.L. region of the CP-violating quantities 
  $\sin 2\alpha$ and $\sin 2\beta$ in the SM, from a fit to the data given in
  Table \protect{\ref{datatable}}.
(From Ref.~\protect\cite{AL-99}.)}
\label{alphabetacorrsm}
\end{figure}

\begin{figure}
\vskip -1.0truein
\centerline{\epsfxsize 3.5 truein \epsfbox {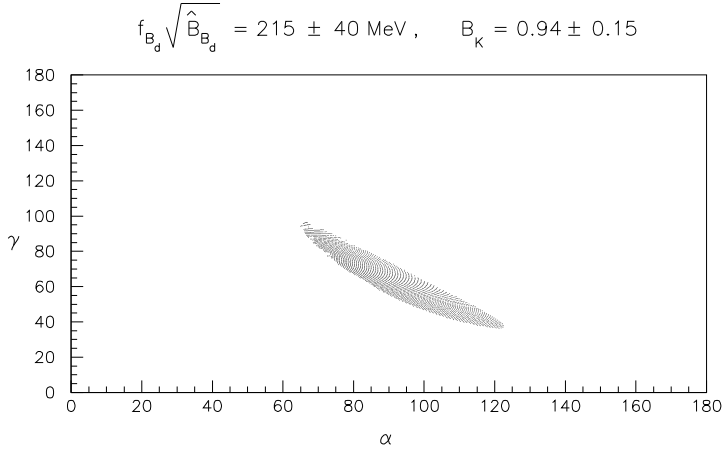}}
\vskip -1.4truein
\caption{\it Allowed 95\% C.L. region of the CP-violating quantities 
  $\alpha$ and $\gamma$ in the SM, from a fit to the data given in Table
  \protect{\ref{datatable}}.
(From Ref.~\protect\cite{AL-99}.)}
\label{alphagammacorrsm}
\end{figure}

Of course, the values of $\alpha$, $\beta$ and $\gamma$ are
correlated, i.e.\ they are not all allowed simultaneously. 
 We illustrate these
correlations in Figs.~\ref{alphabetacorrsm} and \ref{alphagammacorrsm}.
Fig.~\ref{alphabetacorrsm} shows the allowed
region in $\sin 2\alpha$--$\sin 2\beta$ space allowed by the data. And
Fig.~\ref{alphagammacorrsm} shows the allowed (correlated) values of the
CP angles $\alpha$ and $\gamma$. This correlation is roughly linear,
due to the relatively small allowed range of $\beta$
(Eq.~(\ref{CPangleregion})).

% This is Figure 3

% This is Figure 4
\subsection{CKM Fits in Superweak Theories}
As we mentioned earlier, superweak theories of CP violation are now ruled out
by the measurements of the ratio $\epsilon^\prime/\epsilon$ 
\cite{KTEV99,NA31}. We show in this section that  
 a non-trivial constraint on
the CKM phase $\eta$ also results from the present data leaving out the 
information on $\abseps$ (we have not included the measurement of 
 $\epsilon^\prime/\epsilon$ in the analysis either, as discussed in the 
context of our SM-based fits presented earlier). The input parameters 
for this fit are given in Table 1, except that now we leave out $\abseps$ and
$\hat{B}_K$ from the analysis. Thus, we have one data input less compared to 
the SM-fit and only nine parameters to fit (compared to ten in the SM-case).

\begin{figure}
\vskip  -2.0truein
\centerline{\epsfxsize 8.0 truein \epsfbox {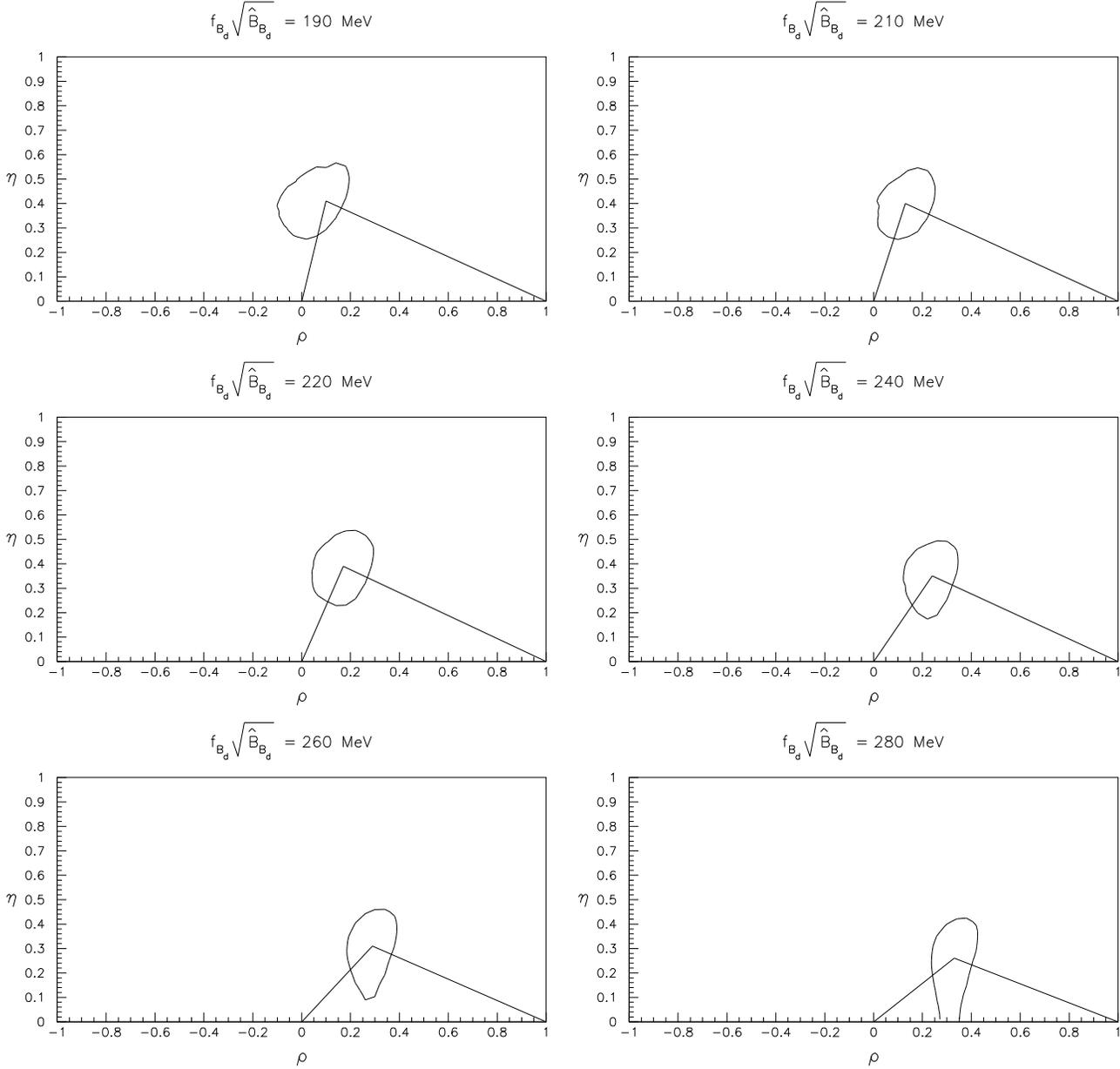}}
\vskip -2.0truein
\caption{\it Allowed regions in $\rho$--$\eta$ space in the Superweak 
theories obtained by leaving out the constraint from $\abseps$, and 
performing a fit to the remaining parameters given in 
Table \protect{\ref{datatable}}. The limit on $\Delta M_s$ is included
  using the amplitude method \protect\cite{Moser97}. The input values
  for $\fbd\protect\sqrt{\hat{B}_{B_d}}$ are shown on top of the individual
 figures. The solid lines
  represent the region with $\chi^2=\chi_{min}^2+6$ corresponding to
  the 95\% C.L.\ region. The triangles show the best fits.}
\label{superweaktriangles1}
\end{figure}

\begin{figure}
\vskip  -2.0truein
\centerline{\epsfxsize 8.0 truein \epsfbox {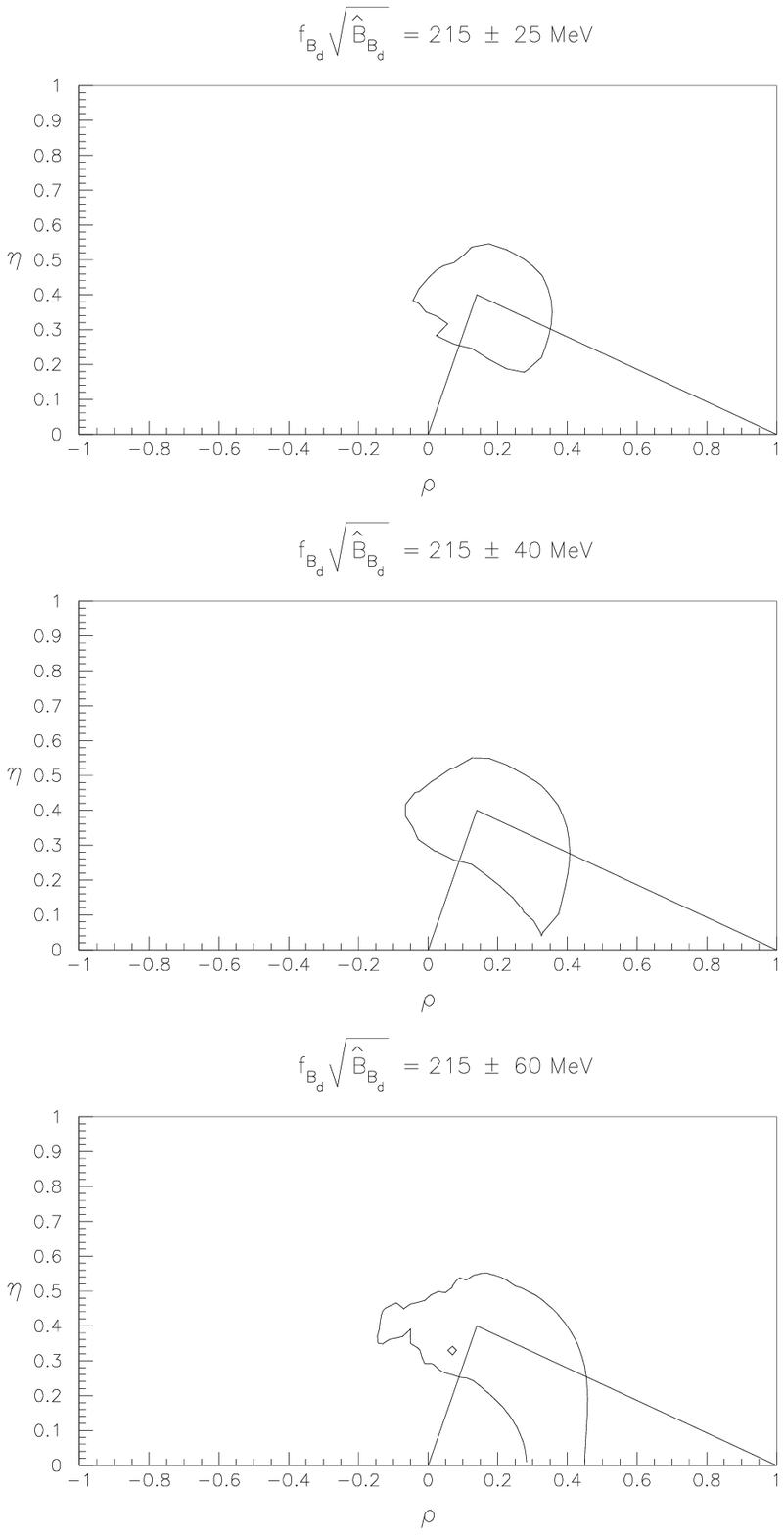}}
\vskip -2.0truein
\caption{\it Allowed regions in $\rho$--$\eta$ space in the Superweak 
theories obtained by leaving out the constraint from $\abseps$, and 
performing a 
  fit to the remaining parameters given in
Table \protect{\ref{datatable}}, assuming that theoretical errors are
Gaussian-distributed. The limit on $\Delta M_s$ is included
using the amplitude method \protect\cite{Moser97}. The input values
for $\fbd\protect\sqrt{\hat{B}_{B_d}}$ used in the fits are shown on 
top of the individual figures. The solid lines
represent the region with $\chi^2=\chi_{min}^2+6$ corresponding to
the 95\% C.L.\ region. The triangles show the best fits.}
\label{superweaktriangles2}
\end{figure}

The most sensitive theoretical parameter in the fits is now 
$\fbd\sqrt{\hat{B}_{B_d}}$. To 
show  the dependence of the allowed CKM-parameter space on this quantity,
we fix its value in performing the fits and vary it in the range $170
~\mbox{MeV} \leq \fbd\sqrt{\hat{B}_{B_d}} \leq ~280 ~\mbox{MeV}$.
 The results for the
allowed 95\% C.L. contour are shown in Fig.~\ref{superweaktriangles1}
for the six values $\fbd\sqrt{\hat{B}_{B_d}}=190, ~210, ~220, ~240, ~260$ and
280 MeV.
 The resulting unitarity triangle for the choice 
$\fbd\sqrt{\hat{B}_{B_d}}=170 ~\mbox{MeV}$ is 
very similar to one shown for  $\fbd\sqrt{\hat{B}_{B_d}}=190 ~\mbox{MeV}$ 
and hence we do not display it. The triangle drawn in each case is the 
best-fit solution. From 
these figures we see that the case $\eta=0$ (superweak model) is ruled 
out for all values of $\fbd\sqrt{\hat{B}_{B_d}}$ in the Lattice-QCD 
range $\fbd\sqrt{\hat{B}_{B_d}} = 215\pm 40 ~\mbox{MeV}$.
Only for very high values of $\fbd\sqrt{\hat{B}_{B_d}}$,
illustrated here by the case $\fbd\sqrt{\hat{B}_{B_d}} = 280~\mbox{MeV}$,
is the superweak theory still compatible with data. Restricting
to the range $190 ~\mbox{MeV} \leq \fbd\sqrt{\hat{B}_{B_d}} \leq ~240 
~\mbox{MeV}$, given by the upper four plots in 
Fig.~\ref{superweaktriangles1}, we see  that at 95\% C.L. the CKM-phase 
$\eta$ is determined to 
lie in the range $0.20 \leq \eta \leq 0.55$. This can also be seen in 
Fig.~\ref{superweaktriangles2} (uppermost of the three curves), where we 
show the resulting 95\% C.L. 
allowed contour using $\fbd\sqrt{\hat{B}_{B_d}}=215\pm 25 ~\mbox{MeV}$,
but assuming that the errors are Gaussian distributed. A comparison of the
allowed $(\rho$ - $\eta)$-contours in Figs.~\ref{superweaktriangles1} and
\ref{superweaktriangles2} also shows that the specific distribution 
assumed for the theoretical error is not crucial. Thus, with the input 
$\fbd\sqrt{\hat{B}_{B_d}}=215\pm 25 ~\mbox{MeV}$, present data
predict a value of $\eta$ well within a factor 3. 
  However, the assumption on the
error of $\fbd\sqrt{\hat{B}_{B_d}}$ does
play a significant role in determining the allowed range of $\eta$. 
For example, using $\fbd\sqrt{\hat{B}_{B_d}}=215\pm 40 
~\mbox{MeV}$ as input, the 95\% C.L. allowed contour comes close to the 
$\rho$-axis,
making the superweak value $\eta=0$ just barely incompatible with data.
This is shown in Fig.~\ref{superweaktriangles2} (second of the three curves
shown here). The superweak theory becomes
compatible with the available data if the theoretical error
on $\fbd\sqrt{\hat{B}_{B_d}}$ is further 
increased to $\pm 60$ MeV. The resulting contour for
$\fbd\sqrt{\hat{B}_{B_d}}=215\pm 60 ~\mbox{MeV}$ is
displayed in Fig.~\ref{superweaktriangles2} (lowest of the three 
curves). 

  We conclude that with the present theoretical knowledge 
$\fbd\sqrt{\hat{B}_{B_d}}=215\pm 40 ~\mbox{MeV}$, the superweak case is
ruled out at 95\% C.L. from the CKM fits, though the value of $\eta$ is not 
determined precisely. 

%
%%%%%%%%%%%%%%%%%%%%%%%%%%%%%%%%%%%%%%%%%%%%%%%%%%%%%%%%%%%%%%%%%%%%%%%%
%
% This is Figure 5
%
%
% This is Figure 6

%%%%%%%%%%%%%%%%%%%%%%%%%%%%%%%%%%%%%%%%%%%%%%%%%%%%%%%%%%%%%%%%%%%%%%%%%%

\section{Unitarity Triangle: A SUSY Profile}

In this section we examine the profile of the unitarity triangle in
supersymmetric (SUSY) theories. The most general
models contain a number of unconstrained phases and so are not
sufficiently predictive to perform such an analysis. However, there is
a class of SUSY models in which these phases are constrained to be
approximately zero, which greatly increases the predictivity.
 In the following
subsections, we discuss aspects of more general SUSY theories, as well
as the details of that class of theories whose effects on the
unitarity triangle can be directly analyzed.

\subsection{Flavour Violation in SUSY Models - Overview}
We begin with a brief review of flavour violation in the minimal
supersymmetric standard model (MSSM).

The low energy effective theory in the MSSM can be specified in terms
of the chiral superfields for the three generations of quarks ($Q_i$,
$U_i^c$, and $D_i^c$) and leptons ($L_i$ and $E_i^c$), chiral
superfields for two Higgs doublets ($H_1$ and $H_2$), and vector
superfields for the gauge group $SU(3)_C \times SU(2)_I \times U(1)_Y$
\cite{Nilles84}. The superpotential is given by
\beq 
W_{\sss MSSM} = f_D^{ij} Q_i D_j H_1 + f_U^{ij} Q_i U_j H_2 +
                  f_L^{ij} E_i L_j H_1 + \mu H_1 H_2.
\label{superpotential}
\eeq
The indices $i,j=1,2,3$ are generation indices and $f_D^{ij}$,
$f_U^{ij}$, $f_L^{ij}$ are Yukawa coupling matrices in the generation
space. A general form of the soft SUSY-breaking term is given by
\begin{eqnarray}
-{\it L}_{\mbox{soft}} &=& (m_Q^2)^i_j \tilde{q}_i \tilde{q}^{\dagger j}
+ (m_D^2)^i_j \tilde{d}_i \tilde{d}^{\dagger j}
+ (m_U^2)^i_j \tilde{u}_i \tilde{u}^{\dagger j}
+ (m_E^2)^i_j \tilde{e}_i \tilde{e}^{\dagger j} \nonumber\\
&+& (m_L^2)^i_j \tilde{\ell}_i \tilde{\ell}^{\dagger j} 
+ \Delta_1^2 h_1^\dagger h_1 + \Delta_2^2 h_2^\dagger h_2
-(B\mu h_1 h_2 + h.c.) \nonumber\\
&+&(A_D^{ij} \tilde{q}_i \tilde{d}_j h_1 + A_U^{ij} \tilde{q}_i \tilde{u}_j 
h_2 + A_L^{ij} \tilde{e}_i \tilde{\ell}_j h_1 + h.c.) \nonumber\\
&+& (\frac{M_1}{2}\tilde{B}\tilde{B} + \frac{M_2}{2}\tilde{W}\tilde{W}
+\frac{M_3}{2}\tilde{G}\tilde{G} + h.c.),
\label{softsusy}
\end{eqnarray}
where $\tilde{q}_i$, $\tilde{u}_i$, $\tilde{d}_i$, $\tilde{\ell}_i$,
$\tilde{e}_i$, $h_1$ and $h_2$ are scalar components of the superfields
$Q_i$, $U_i$, $D_i$, $L_i$, $E_i$, $H_1$ and $H_2$, respectively, and
$\tilde{B}$, $\tilde{W}$ and $\tilde{G}$ are the $U(1)$, $SU(2)$ and $SU(3)$
gauge fermions, respectively. The SUSY-breaking parameters $(m_{F})^i_j$,
with $m_F=m_D,m_U,m_L,m_E$ and the trilinear scalar couplings $A_D^{ij},
A_U^{ij}$ and $A_L^{ij}$ are $3 \times 3$ matrices in the flavour space.
It is obvious that supersymmetric theories have an incredibly complicated 
flavour structure, resulting in a large number of {\it a priori} unknown
mixing angles, which cannot be determined theoretically. Present
measurements and limits on the FCNC processes do provide some constraints
on these mixing angles \cite{ML97}. We shall not follow this route 
here.  

Alternatively, one could put restrictions on the soft SUSY-breaking
(SSB) terms. The ones most discussed in the literature are those which
find their rationale in supergravity (SUGRA) models, in which it is
assumed that the SSB terms have universal structures at the Planck
scale, following from the assumption that the hidden sector of $N=1$
SUGRA theory is flavour-blind. This results in the universal scalar
mass, $m_0$, with $(m_Q^2)^i_j =(m_E^2)^i_j= ,...=m_0^2 \delta^i_j$;
$\Delta_1^2=\Delta_2^2=\Delta_0^2$, universal $A$-terms,
$A_D^{ij}=f_D^{ij}Am_0,$ etc., and universal gaugino masses $M_g$,
defined as $M_1=M_2=M_3=M_g$. These universal structures are required
in order to suppress FCNC processes.  The scenario with the additional
constraint $m_0=\Delta_0$ is called the minimal SUGRA model. In other
theoretical scenarios, it is not necessary to invoke universal SSB
terms. In order to make testable predictions it is sufficient to
restrict all flavour violations in the charged-current sector, which
are determined by the known CKM angles \cite{CDGG98}. We shall be
mostly dealing here with this scenario, known as {\it minimal flavour
  violation}, as well as SUGRA-type scenarios.

Even in these restricted scenarios, one is confronted with the complex phases
residing in the $W_{\sss MSSM}$ and $L_{\mbox{soft}}$ part of the supersymmetric
lagrangian.
In general, MSSM models have three physical phases, apart from the QCD
vacuum parameter $\bar{\theta}_{\sss QCD}$ which we shall take to be
zero. The three phases are: (i) the CKM phase represented here by the
Wolfenstein parameter $\eta$, (ii) the phase $\theta_A=\arg (A)$, and
(iii) the phase $\theta_\mu=\arg (\mu)$ \cite{DGH85}. The last two
phases are peculiar to SUSY models and their effects must be taken
into account in a general supersymmetric framework. In particular, the
CP-violating asymmetries which result from the interference between
mixing and decay amplitudes can
produce non-standard effects. Concentrating here on the $\Delta B=2$
amplitudes, two new phases $\theta_d$ and $\theta_s$ arise, which can
be parametrized as follows \cite{effsusy}:
\beq
\theta_{d,s} = \frac{1}{2} \arg \left(\frac{\langle B_{d,s} \vert {\it 
H}_{eff}^{\sss SUSY} \vert \bar{B}_{d,s} \rangle}{\langle B_{d,s} \vert 
{\it H}_{eff}^{\sss SM} \vert \bar{B}_{d,s} \rangle} \right) ~,
\eeq
where ${\it H}^{\sss SUSY}$ is the effective Hamiltonian including
both the SM degrees of freedom and the SUSY contributions. Thus,
CP-violating asymmetries in $B$ decays would involve not only the
phases $\alpha$, $\beta$ and $\gamma$, defined previously, but
additionally $\theta_d$ or $\theta_s$. In other words, the SUSY
contributions to the real parts of $M_{12}(B_d)$ and $M_{12}(B_s)$ are
{\it no longer proportional} to the CKM matrix elements $V_{td}
V_{tb}^*$ and $V_{ts}V_{tb}^*$, respectively. If $\theta_d$ or
$\theta_s$ were unconstrained, one could not make firm predictions
about the CP asymmetries in SUSY models. In this case, an analysis of
the profile of the unitarity triangle in such models would be futile.

However, the experimental upper limits on the electric dipole moments
(EDM) of the neutron and electron \cite{PDG98} do provide constraints
on the phases $\theta_\mu$ and $\theta_A$ \cite{FOS95}. In SUGRA
models with {\it a priori} complex parameters $A$ and $\mu$, the phase
$\theta_\mu$ is strongly bounded with $\theta_\mu < 0.01 \pi$
\cite{Nihei97}. The phase $\theta_A$ can be of $O(1)$ in the small
$\theta_\mu$ region, as far as the EDMs are concerned. In both the
$\Delta S=2$ and $\Delta B=2$ transitions, and for low-to-moderate
values of $\tan \upsilon$ \footnote{In supersymmetric jargon, the
  quantity $\tan \beta$ is used to define the ratio of the two vacuum
  expectation values (vevs) $\tan \beta \equiv v_u/v_d$, where
  $v_d(v_u)$ is the vev of the Higgs field which couples exclusively
  to down-type (up-type) quarks and leptons. (See, for example, the
  review by Haber in Ref.~\cite{PDG98}). However, in discussing
  flavour physics, the symbol $\beta$ is traditionally reserved for
  one of the angles of the unitarity triangle. To avoid confusion, we
  will call the ratio of the vevs $\tan \upsilon$.}, it has been shown
that $\theta_A$ does not change the phase of either the matrix element
$M_{12}(K)$ \cite{DGH85} or of $M_{12}(B)$ \cite{Nihei97}. Hence, in
SUGRA models, $\arg M_{12}(B)|_{\sss SUGRA} = \arg M_{12}(B)|_{\sss
  SM}=\arg (\xi_t^2)$, where $\xi_t=V_{td}^*V_{tb}$. Likewise, the
phase of the SUSY contribution in $M_{12}(K)$ is aligned with the
phase of the $t\bar{t}$-contribution in $M_{12}(K)$, given by $\arg
(V_{td}V_{ts}^*)$. Thus, in these models, one can effectively set
$\theta_d \simeq 0$ and $\theta_s \simeq 0$, so that the CP-violating
asymmetries give information about the SM phases $\alpha$, $\beta$ and
$\gamma$. Hence, an analysis of the UT and CP-violating phases
$\alpha$, $\beta$ and $\gamma$ can be carried out in a very similar
fashion as in the SM, taking into account the additional contributions
to $M_{12}(K)$ and $M_{12}(B)$.

For large-$\tan\upsilon$ solutions, one has to extend the basis of
$H_{eff} (\Delta B=2)$ so as to include new operators whose contribution 
is small in the low-$\tan\upsilon$ limit. The resulting effective Hamiltonian
is given by
\beq 
H_{eff}(\Delta B=2) = \frac{G_F^2M_W^2}{2\pi^2} \sum_{i=1}^{3} 
C_i(\mu)O_{i}~,
\label{SMeffham}
\eeq
where $O_1=\bar{d}_L^\alpha\gamma_\mu b_L^\alpha \bar{d}_L^\beta
\gamma^\mu b_L^\beta$, $O_2= \bar{d}_L^\alpha b_R^\alpha
\bar{d}_L^\beta b_R^\beta$ and $O_3= \bar{d}_L^\alpha b_R^\beta
\bar{d}_L^\beta b_R^\alpha$ and $C_i$ are the Wilson coefficients
\cite{Brancoetal,CS98}. The coefficients $C_1(\mu)$ and $C_2(\mu)$ are
real relative to the SM contribution. However, the chargino
contributions to $C_3(\mu)$ are generally complex relative to the SM
contribution and can generate a new phase shift in the
$B^0$--$\overline{B^0}$ mixing amplitude \cite{DMV98,BK98-1}. This
effect is in fact significant for large $\tan \upsilon$
\cite{Brancoetal}, since $C_3(\mu)$ is proportional to $(m_b/m_W\cos
\beta)^2$. How large this additional phase $(\theta_d $ and
$\theta_s$) can be depends on how the constraints from EDM are
imposed. For example, Baek and Ko \cite{BK98-1} find that in the MSSM
without imposing the EDM constraint, one has $2 \vert \theta_d \vert
\leq 6^\circ$ for a light stop and large $\tan \upsilon$ but this phase
becomes practically zero if the EDM constraints \cite{CKP98} are
imposed.

In view of the foregoing, we shall restrict ourselves to a class of
SUSY models in which the following features, related to flavour
mixing, hold:
\begin{itemize}
\item The squark flavour mixing matrix which diagonalizes the squark
mass matrix is approximately the same as the corresponding quark
mixing matrix $V_{CKM}$, apart from the left-right mixing of the top
squarks.

\item The phases $\theta_d$ and $\theta_s$ are negligible in the entire
$\tan \upsilon$ plane, once the constraints from the EDMs of neutron and
lepton are consistently  imposed.

\item The first- and second-generation squarks with the same quantum
  numbers remain highly degenerate in masses but the third-generation
  squarks, especially the top squark, can be significantly lighter due
  to the renormalization effect of the top Yukawa coupling constants.
\end{itemize}
These features lead to an enormous simplification in the flavour
structure of the SUSY contributions to flavour-changing processes. In
particular, SUSY contributions to the transitions $b \to s$, $b \to d$
and $s \to d$ are proportional to the CKM factors, $V_{tb}V_{ts}^*$,
$V_{tb}V_{td}^*$ and $V_{ts}V_{td}^*$, respectively. Similarly, the
SUSY contributions to the mass differences $M_{12}(B_s)$,  
$M_{12}(B_d)$ and $M_{12}(K)$ are proportional to the CKM factors
$(V_{tb}V_{ts}^*)^2$, $(V_{tb}V_{td}^*)^2$ and $(V_{ts}V_{td}^*)^2$,
respectively. These are precisely the same factors which govern the
contribution of the top quark in these transitions in the standard
model. Thus, the supersymmetric contributions can be implemented in a
straightforward way by adding a (supersymmetric) piece in each of the
above mentioned amplitudes to the corresponding top quark contribution
in the SM.

\subsection{NLO Corrections to $\delmd$, $\delms$ and $\epsilon$ in Minimal 
SUSY Flavour Violation}

A number of SUSY models share the features mentioned in the previous
subsection,
%foremost among them the SUGRA models, 
and the supersymmetric contributions to the mass differences
$M_{12}(B)$ and $M_{12}(K)$ have been analyzed in a number of papers
\cite{Nihei97,Brancoetal,Gotoetal96,Gotoetal97,Gotoetal98-1,Gotoetal98-2},
following the pioneering work of Ref.~\cite{BBMR91}. Following these
papers, $\delmd$ can be expressed as:
\begin{eqnarray}
\delmd & = & \frac{G_F^2}{6\pi^2}M_W^2M_B\left(\fbb\right)\hat{\eta}_B
\times \nonumber \\
&& ~~~~~~~~~~~~~~~~~~~~
\left[A_{SM}(B) + A_{H^\pm}(B) + A_{\chi^\pm}(B) + A_{\tilde{g}}(B) \right]~,
\label{delmdsusy}
\end{eqnarray}
where the function $A_{SM}(B)$ can be written by inspection from
Eq.~(\ref{bdmixing}):
\beq
A_{SM}(B)= y_t f_2(y_t) \vert V_{td}^* V_{tb} \vert^2 ~.
\eeq
The expressions for $A_{H^\pm}(B)$, $A_{\chi^\pm}(B)$ and
$A_{\tilde{g}}(B)$ are obtained from the SUSY box diagrams.
Here, $H^\pm$, $\chi^\pm_j$, $\tilde{t}_a$ and $\tilde{d}_i$ represent,
respectively, the charged Higgs, chargino, stop and down-type
squarks. The contribution of the intermediate states involving
neutralinos is small and usually neglected. The expressions for
$A_{H^\pm}(B)$, $A_{\chi^\pm}(B)$ and $A_{\tilde{g}}(B)$ are given
explicitly in the literature \cite{Brancoetal,Gotoetal96,BBMR91}.

We shall not be using the measured value of the mass difference $\Delta
M_K$ due to the uncertain contribution of the long-distance
contribution. However, $\abseps$ is a short-distance dominated
quantity and in supersymmetric theories can be expressed as follows:
\begin{eqnarray}
\abseps & = & \frac{G_F^2f_K^2M_KM_W^2}{6\sqrt{2}\pi^2\Delta M_K}
\hat{B}_K
\left[\mbox{Im}~A_{SM}(K) + \mbox{Im}~A_{H^\pm}(K) \right. \nonumber \\
&& ~~~~~~~~~~~~~~~~~~~~~~~~~~~~~~
\left. + \mbox{Im}~A_{\chi^\pm}(K) + \mbox{Im}~A_{\tilde{g}}(K) \right]~,
\label{epsilonsusy}
\end{eqnarray}
where, again by inspection with the SM expression for $\abseps$ given 
in Eq.~(\ref{eps}), one has   
\beq
\mbox{Im}~A_{SM}(K) =
A^2\lambda^6\eta
\bigl(y_c\left\{\hat{\eta}_{ct}f_3(y_c,y_t)-\hat{\eta}_{cc}\right\}
+ \hat{\eta}_{tt}y_tf_2(y_t)A^2\lambda^4(1-\rho)\bigr).
\label{asm}
\eeq
The expressions for $\mbox{Im}~A_{H^\pm}(K)$,
$\mbox{Im}~A_{\chi^\pm}(B)$ and $\mbox{Im}~A_{\tilde{g}}(B)$ can be
found in Refs.~\cite{Brancoetal,Gotoetal96,BBMR91}.

For the analysis reported here, we follow the scenario called {\it
  minimal flavour violation} in Ref.~\cite{CDGG98}. In this class of
supersymmetric theories, apart from the SM degrees of freedom, only
charged Higgses, charginos and a light stop (assumed to be
right-handed) contribute, with all other supersymmetric particles
integrated out. This scenario is effectively implemented in a class of
SUGRA models (both minimal and non-minimal) and gauge-mediated models
\cite{Dine93}, in which the first two squark generations are heavy and
the contribution from the intermediate gluino-squark states is small
\cite{Nihei97,Gotoetal96,Gotoetal97,Gotoetal98-1,Gotoetal98-2}.

For these models, the next-to-leading-order (NLO) corrections for
$\delmd$, $\delms$ and $\abseps$ can be found in Ref.~\cite{KS98}.
Moreover, the branching ratio ${\it B}(B \to X_s \gamma)$ has been
calculated in Ref.~\cite{CDGG98}. We make use of this information and
quantitatively examine the unitarity triangle, CP-violating
asymmetries and their correlations for this class of supersymmetric
theories. The phenomenological profiles of the unitarity triangle and
CP phases for the SM and this class of supersymmetric models can thus
be meaningfully compared. Given the high precision on the phases
$\alpha$, $\beta$ and $\gamma$ expected from experiments at
$B$-factories and hadron colliders, a quantitative comparison of this
kind could provide a means of discriminating between the SM and this
class of MSSM's.

The NLO QCD-corrected effective Hamiltonian
for $\Delta B=2$ transitions in the minimal flavour violation SUSY
framework can be expressed as follows \cite{KS98}:
\beq
H_{eff} = \frac{G_F^2}{4 \pi^2} (V_{td}V_{tb}^*)^2 \hat{\eta}_{2,S}(B) S 
O_{LL}~,
\label{effHsusy}
\eeq 
where the NLO QCD correction factor $\hat{\eta}_{2,S}(B)$ is given by
\cite{KS98}:
\beq
\hat{\eta}_{2,S}(B) = \alpha_s(m_W) \gamma^{(0)}/(2 \beta_{n_f}^{(0)}) \left[
1 + \frac{\alpha_s(m_W)}{4 \pi} \left(\frac{D}{S} + Z_{n_f}\right) \right]~,
\label{eta2s}
\eeq
in which $n_f$ is the number of active quark flavours (here $n_f=5$), the
quantity $Z_{n_f}$ is defined below, 
and $\gamma^{(0)}$ and $\beta_{n_f}^{(0)}$ are the lowest order perturbative
QCD $\beta$-function and the anomalous dimension, respectively.
The operator $O_{LL}=O_1$ is the one which is
present in the SM, previously defined in the discussion following
Eq.~(\ref{SMeffham}).
The explicit expression for the function $S$ can be obtained from
Ref.~\cite{BBMR91} and for $D$ it is given in Ref.~\cite{KS98}, where
it is derived in the NDR (naive dimensional regularization) scheme
using $\overline{MS}$-renormalization.

The Hamiltonian given above for $B_d^0$--$\overline{B_d^0}$ mixing
leads to the mass difference
\beq
\delmd = \frac{G_F^2}{6 \pi^2} (V_{td}V_{tb}^*)^2 \hat{\eta}_{2,S}(B) S
(\fbd^2 \hat{B}_{B_d})~.
\label{deltamdsusy}
\eeq
The corresponding expression for $\delms$ is obtained by making the
appropriate replacements. Since the QCD correction factors are
identical for $\delmd$ and $\delms$, it follows that the quantities
$\delmd$ and $\delms$ are enhanced by the same factor in minimal
flavour violation supersymmetry, as compared to their SM values, but
the ratio $\delms/\delmd$ in this theory is the same as in the SM.

The NLO QCD-corrected Hamiltonian for $\Delta S=2$ transitions in the
minimal flavour violation supersymmetric framework has also been
obtained in Ref.~\cite{KS98}. From this, the result for $\epsilon$ can
be written as:
\begin{eqnarray}
\abseps &=& \frac{G_F^2f_K^2M_KM_W^2}{6\sqrt{2}\pi^2\Delta M_K}
\hat{B}_K\left(A^2\lambda^6\eta\right)
\bigl(y_c\left\{\hat{\eta}_{ct}f_3(y_c,y_t)-\hat{\eta}_{cc}\right\}
 \nonumber \\
&~& ~~~~~~~~~~~~~~~~~~~~+
~\hat{\eta}_{2}(K)SA^2\lambda^4(1-\rho)\bigr),
\label{eps2}
\end{eqnarray}
where the NLO QCD correction factor is \cite{KS98}:
\begin{eqnarray}
\hat{\eta}_{2}(K) &=& \alpha_s(m_c)^{\gamma^{(0)}/(2\beta_3^{(0)})}
\left(\frac{\alpha_s(m_b)}{\alpha_s(m_c)}\right)^{\gamma^{(0)}/(2\beta_4^{(0)})}
\left(\frac{\alpha_s(M_W)}{\alpha_s(m_b)}\right)^{\gamma^{(0)}/(2\beta_5^{(0)})} 
\times
\nonumber\\
&& \hskip-12truemm
\left[ 1 + \frac{\alpha_s(m_c)}{4 \pi} (Z_3 - Z_4) +
\frac{\alpha_s(m_b)}{4 \pi} (Z_4 - Z_5) +
\frac{\alpha_s(M_W)}{4 \pi} (\frac{D}{S}+ Z_5) \right] ~.
\label{eta2k}
\end{eqnarray}
Here
\beq
Z_{n_f} = \frac{\gamma_{n_f}^{(1)}}{2 \beta_{n_f}^{(0)}} -
\frac{\gamma^{(0)}}{2 {\beta_{n_f}^{(0)}}^2} \beta_{n_f}^{(1)}~,
\eeq
and the quantities entering in Eqs.~(\ref{eta2s}) and (\ref{eta2k})
are the coefficients of the well-known beta function and anomalous
dimensions in QCD:
 The ratio
\beq 
\frac{\hat{\eta}_{2,S}(B)(NLO)}{\hat{\eta}_{2,S}(B)(LO)}
= 1 +\frac{\alpha_s(M_W)}{4\pi}
(\frac{D}{S} + Z_5),
\label{eta2lnl}
\eeq 
is worked out numerically in Ref.~\cite{KS98} as a function of the
supersymmetric parameters (chargino mass $m_{\chi_2}$, mass of the
lighter of the two stops $m_{\tilde{t}_R}$, and the mixing angle
$\phi$ in the stop sector). This ratio is remarkably stable against
variations in the mentioned parameters and is found numerically to be
about $0.89$. Since in the LO approximation the QCD correction factor
${\hat \eta}_{2,S}(B)(LO)$ is the same in the SM and SUSY,
the QCD correction factor ${\hat \eta}_{2,S}(B)(NLO)$ entering in the
expressions for $\Delta M_d$ and $\Delta M_s$ in the MSSM is found to
be ${\hat \eta}_{2,S}(B)(NLO) = 0.51$ in the $\overline{MS}$-scheme. This is 
to be compared with the
corresponding quantity ${\hat \eta}_B = 0.55$ in the SM. Thus, NLO
corrections in $\Delta M_d$ (and $\Delta M_s$) are similar in the SM
and MSSM, but not identical.

The expression for ${\hat \eta}_{2,S}(K)(NLO)/{\hat
\eta}_{2,S}(K)(LO)$ can be expressed in terms of the ratio ${\hat
\eta}_{2,S}(B)(NLO)/{\hat \eta}_{2,S}(B)(LO)$ given above and the
flavour-dependent matching factors $Z_{n_f}$:
\begin{eqnarray}
\frac{\hat{\eta}_{2,S}(K)(NLO)}{\hat{\eta}_{2,S}(K)(LO)}
& = & \frac{\hat{\eta}_{2,S}(B)(NLO)}{\hat{\eta}_{2,S}(B)(LO)} 
+ {\alpha_s(m_c) \over 4\pi} (Z_3 - Z_4)
+ {\alpha_s(m_b) \over 4\pi} (Z_4 - Z_5) \nonumber\\
& \simeq & 0.884 ~,
\end{eqnarray}
where we have used the numerical value $\hat{\eta}_{2,S}(B)(NLO) /
\hat{\eta}_{2,S}(B)(LO) = 0.89$ calculated by Krauss and Soff
\cite{KS98}, along with $\alpha_s(m_c) = 0.34$ and $\alpha_s(m_b) =
0.22$. Using the expression for the quantity ${\hat
  \eta}_{2,S}(K)(LO)$, which is given by the prefactor multiplying the
square bracket in Eq.~(\ref{eta2k}), one gets ${\hat
  \eta}_{2,S}(K)(NLO) = 0.53$ in the $\overline{MS}$-scheme. This is
to be compared with the corresponding QCD correction factor in the SM,
${\hat\eta}_{tt} = 0.57$, given in Table \ref{datatable}. Thus the two
NLO factors are again very similar but not identical.

Following the above discussion, the SUSY contributions to $\delmd$,
$\delms$ and $\abseps$ in supersymmetric theories are incorporated in our
analysis in a simple form:
\begin{eqnarray}
\delmd &=& \delmd (SM) [ 1 +
f_d(m_{\chi_2^\pm},m_{\tilde{t}_R},
m_{H^\pm}, \tan \upsilon, \phi) ], \nonumber \\
\delms &=& \delms (SM) [ 1 +
f_s(m_{\chi_2^\pm},m_{\tilde{t}_R},
m_{H^\pm}, \tan \upsilon, \phi) ], \nonumber \\
\abseps &=& \frac{G_F^2f_K^2M_KM_W^2}{6\sqrt{2}\pi^2\Delta M_K}
\hat{B}_K\left(A^2\lambda^6\eta\right)
\bigl(y_c\left\{\hat{\eta}_{ct}f_3(y_c,y_t)-\hat{\eta}_{cc}\right\}
 \nonumber \\
&~& ~~~+
~\hat{\eta}_{tt}y_tf_2(y_t)[1 + f_\epsilon
(m_{\chi_2^\pm},m_{\tilde{t}_2}, m_{H^\pm},
\tan \upsilon, \phi)] A^2\lambda^4(1-\rho)\bigr).
\label{susyformel}
\end{eqnarray}
Here, $\phi$ is the $LR$-mixing angle in the stop sector. The quantities 
$f_d$, $f_s$ and $f_\epsilon$ can be expressed as %
\beq
\label{fis}
f_d = f_s= \frac{\hat{\eta}_{2,S}(B)}{\hat{\eta}_B} R_{\Delta_d}(S) ~,~~
f_\epsilon = \frac{\hat{\eta}_{2,S}(K)}{\hat{\eta}_{tt}}R_{\Delta_d}(S),
\eeq
where $R_{\Delta_d}(S)$ is defined as
\beq
R_{\Delta_d}(S) \equiv { \Delta M_d(SUSY) \over \Delta M_d(SM) } (LO)
= {S \over y_t f_2(y_t)} ~.
\eeq
The functions $f_{i}$, $i=d,s,\epsilon$ are all positive definite, i.e.\
the supersymmetric contributions add {\it constructively} to the SM 
contributions in the entire allowed supersymmetric parameter space. We
find that the two QCD correction factors appearing in Eq.~(\ref{fis})   
are numerically very close to one another, with
$\hat{\eta}_{2,S}(B)/\hat{\eta}_B \simeq
{\hat\eta}_{2,S}(K)/\hat{\eta}_{tt} = 0.93$. Thus, to an excellent
approximation, one has $f_d = f_s = f_\epsilon \equiv f$.

\begin{figure}
\vskip -2.0truein
\centerline{\epsfxsize 8.0 truein \epsfbox {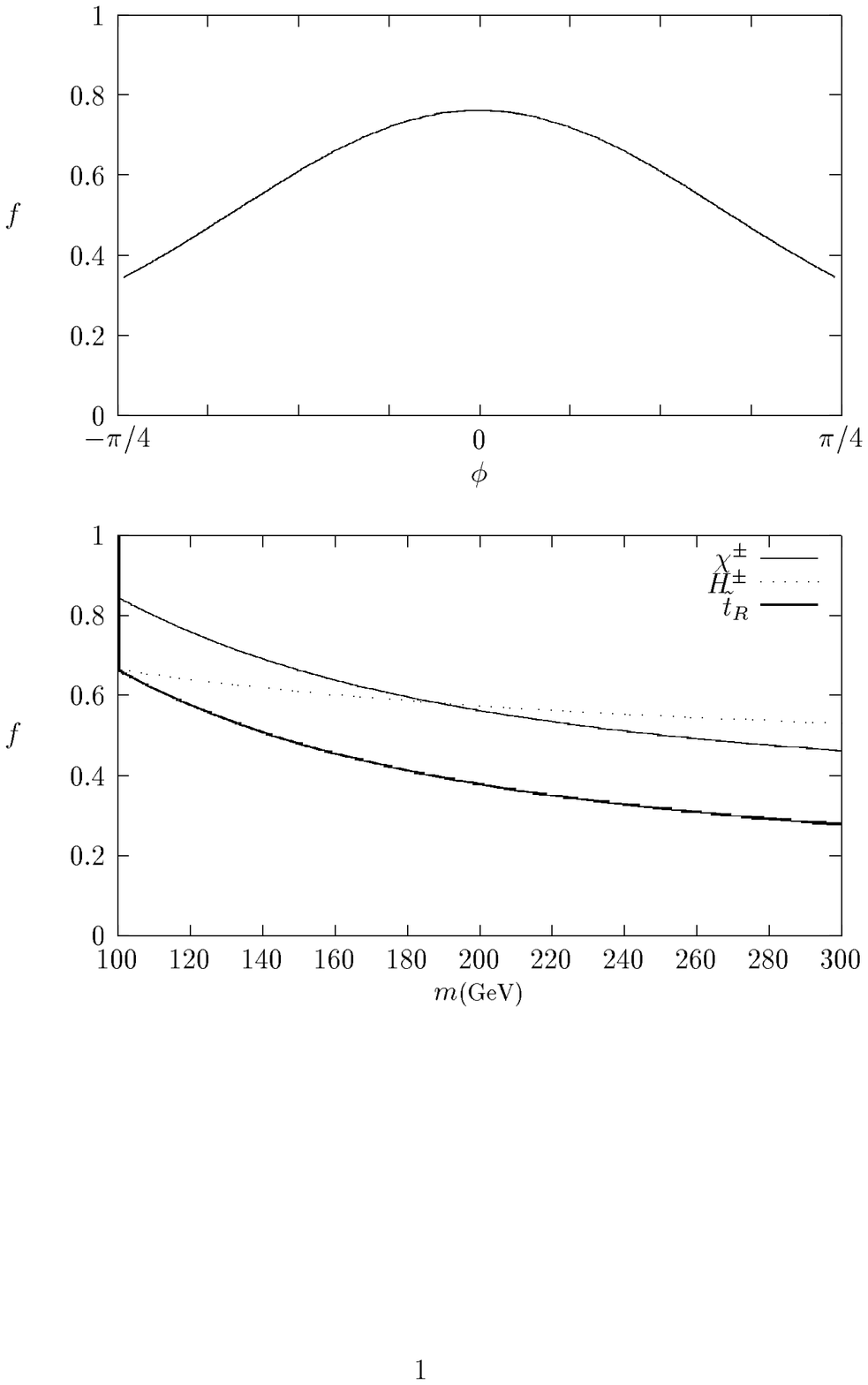}}
\vskip -3.0truein
\caption{\it Dependence of the supersymmetric function $f$ on 
the $LR$-mixing angle in the stop sector, $\phi$ (upper figure), and on  
$m_{\tilde{\chi}^\pm_2}$, $m_{H^\pm}$ and $m_{\tilde{t}_R}$
(lower figure), for $\tan \upsilon=2$; values of 
the other supersymmetric parameters are stated in the text. }
\label{susyfunction2}
\end{figure}

\begin{figure}
\vskip -2.0truein
\centerline{\epsfxsize 8.0 truein \epsfbox {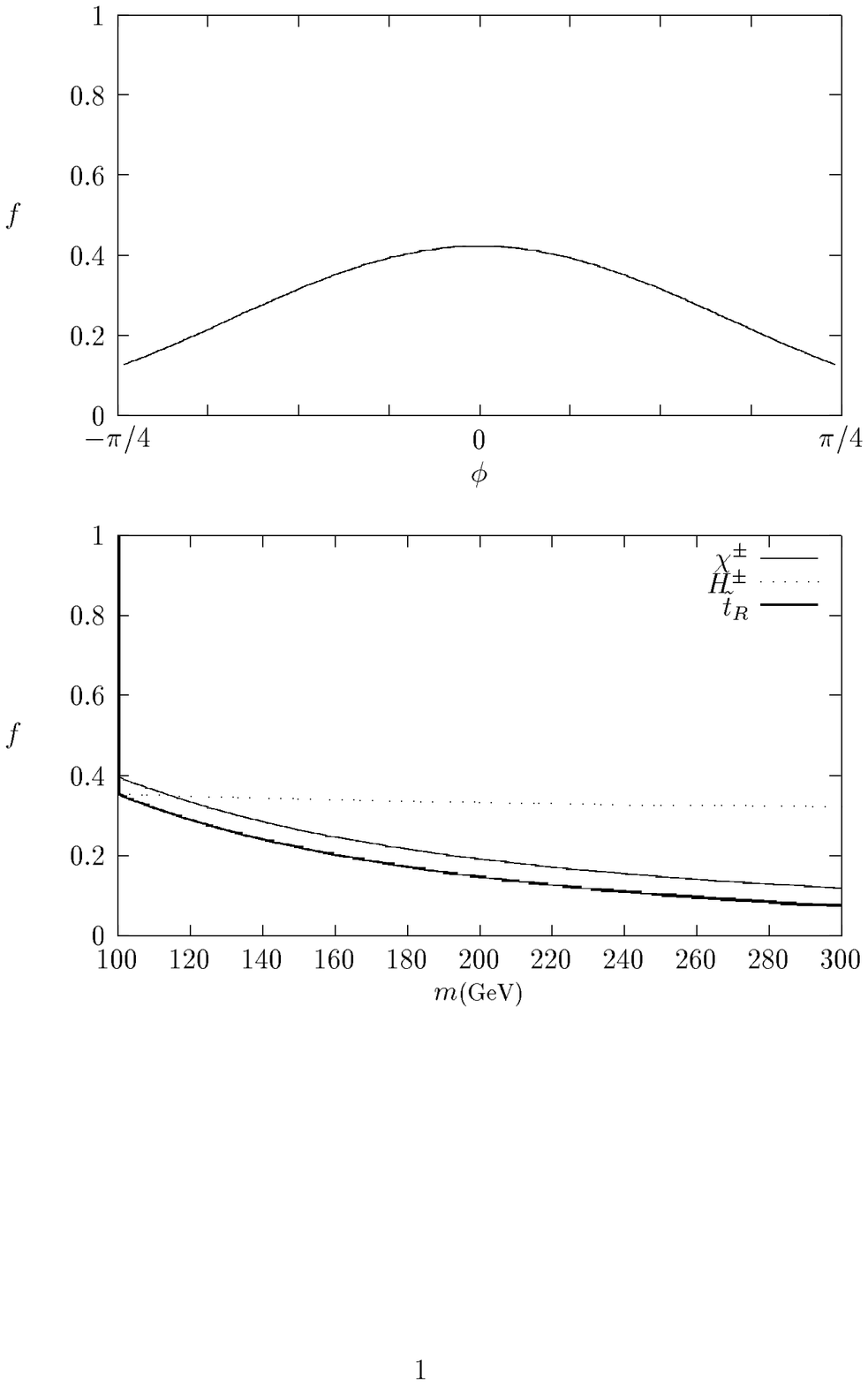}}
\vskip -3.0truein
\caption{\it Dependence of the supersymmetric function $f$ on
the $LR$-mixing angle in the stop sector, $\phi$ (upper figure), and on
$m_{\tilde{\chi}^\pm_2}$, $m_{H^\pm}$ and $m_{\tilde{t}_R}$
(lower figure), for $\tan \upsilon=4$; values of  
the other supersymmetric parameters are stated in the text. }
\label{susyfunction4}
\end{figure}

How big can $f$ be? This quantity is a function of the masses of the
top squark, chargino and the charged Higgs, $m_{\tilde{t}_R}$,
$m_{\tilde{\chi}^\pm_2}$ and $m_{H^\pm}$, respectively, as well as of
$\tan\upsilon$. The maximum allowed value of $f$ depends on the model
(minimal SUGRA, non-minimal SUGRA, MSSM with constraints from EDMs,
etc.).  We have numerically calculated the quantity $f$ by varying the
SUSY parameters $\phi$, $m_{\tilde{t}_R}$, $m_{\chi_2}$, $m_{H^\pm}$
and $\tan \upsilon$.  Using, for the sake of illustration,
$m_{\chi_2^\pm}=m_{\tilde{t}_R}=m_{H^\pm} =100$ GeV,
$m_{\chi_1^\pm}=400$ GeV and $\tan \upsilon=2$, and all other
supersymmetric masses much heavier, of $O(1)$ TeV, we find that the
quantity $f$ varies in the range:
\beq
 0.4 \leq f \leq 0.8 ~~~\mbox{for}~~~ \vert \phi \vert \leq \pi/4 ~,
\label{fnumber}
\eeq
with the maximum value of $f$ being at $\phi=0$. This is shown in 
Fig.~\ref{susyfunction2}, where we have plotted the function $f$ against 
$\phi$ (upper figure), and against $m_{\chi_2^\pm}$, $m_{\tilde{t}_R}$ and 
$m_{H^\pm}$ (lower figure), varying 
one parameter at a time and holding the others fixed to their stated values 
given above. These parametric values are allowed by the constraints 
from the NLO analysis of the decay $B \to X_s + \gamma$ reported in
Ref.~\cite{CDGG98}, as well as from direct searches of the
supersymmetric particles \cite{PDG98}.
The allowed value of $f$ decreases as $m_{\tilde{t}_R}$,
$m_{\chi_2}$ and $m_{H^\pm}$ increase, though the dependence of $f$ on
$m_{H^\pm}$ is rather mild due to the compensating effect of the
$H^\pm$ and chargino contributions in the MSSM, as observed in
Ref.~\cite{CDGG98}. Likewise, the allowed range of
$f$ is reduced as $\tan \upsilon$ increases, as shown in 
Fig.~\ref{susyfunction4} for $\tan \upsilon=4$, in which case one has
$0.15 \leq f \leq 0.42$ for $\vert \phi \vert \leq \pi/4$.
 This sets the size of $f$ allowed by the present 
constraints in the minimal flavour violation version of the MSSM.

If additional constraints on the supersymmetry breaking parameters are
imposed, as is the case in the minimal and non-minimal versions of the
SUGRA models, then the allowed values of $f$ will be further
restricted. A complete NLO analysis of $f$ would require a
monte-carlo approach implementing all the experimental and theoretical
constraints (such as the SUGRA-type mass relations). In particular,
the NLO correlation between ${\it B}(B \to X_s \gamma)$ and $f$ has
to be studied in an analogous fashion, as has been done, for example,
in Refs.~\cite{Gotoetal98-1,Gotoetal98-2} with the leading order SUSY
effects. 

In this paper we adopt an approximate method to constrain $f$ in
SUGRA-type models. We take the maximum allowed values of the quantity
$R_{\Delta_d}(S)$, defined earlier, from the existing LO analysis of
the same and obtain $f$ by using Eq.~(\ref{fis}). For the sake of
definiteness, we use the updated work of Goto et al.\ 
\cite{Gotoetal98-1,Gotoetal98-2}.

 From the published results we conclude that typically 
$f$ can be as large as $0.45$ in non-minimal SUGRA models for low $\tan
\upsilon$ (typically $\tan \upsilon=2$) \cite{Gotoetal98-1}, and
approximately half of this value in minimal SUGRA models
\cite{Nihei97,Gotoetal97,Gotoetal98-1}. Relaxing the SUGRA mass
constraints, admitting complex values of $A$ and $\mu$ but
incorporating the EDM constraints, and imposing the constraints mentioned
above,  Baek and Ko \cite{BK98-1} find that
$f$ could be as large as $f=0.75$. In all cases, the value of $f$
decreases with increasing $\tan \upsilon$ or increasing
$m_{\tilde{\chi}^\pm_2}$ and $m_{\tilde{t}_R}$, as noted above.

\subsection{SUSY Fits}
%
%%%%%%%%%%%%%%%%%%%%%%%%%%%%%%%%%%%%%%%%%%%%%%%%%%%%%%%%%%%%%%%%%%%%%%%%%
% This is Figure 7
%
%
% This is Figure 8

%%%%%%%%%%%%%%%%%%%%%%%%%%%%%%%%%%%%%%%%%%%%%%%%%%%%%%%%%%%%%%%%%%%%%%%%%
% This is Figure 9
\begin{figure}
\vskip -2.0truein
\centerline{\epsfxsize 8.0 truein \epsfbox {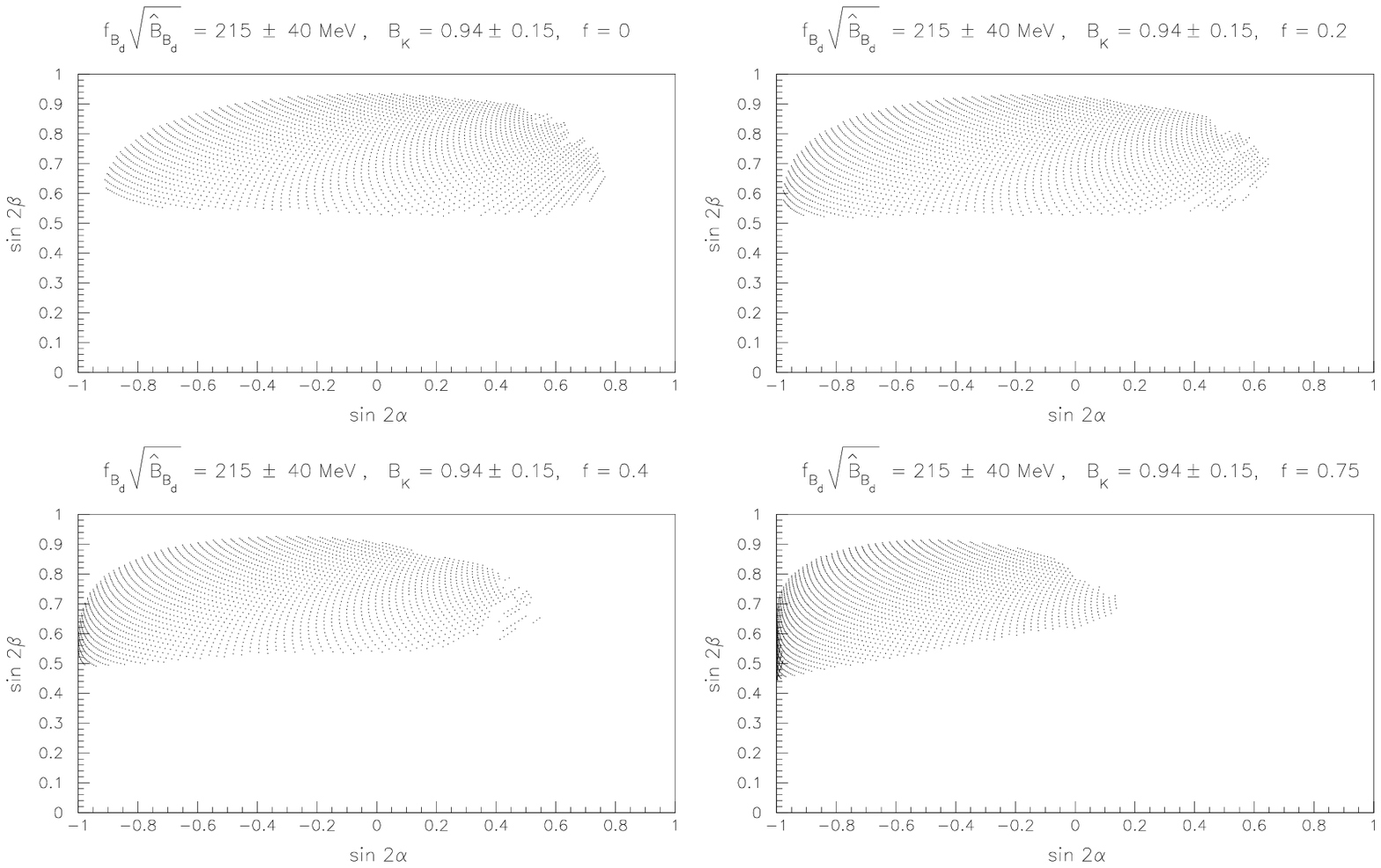}}
\vskip -4.7truein
\caption{\it Allowed 95\% C.L. region of the CP-violating quantities
  $\sin 2\alpha$ and $\sin 2\beta$, from a fit to the data given in
  Table \protect{\ref{datatable}}. The upper left plot ($f=0$)
  corresponds to the SM, while the other plots ($f=0.2$, 0.4, 0.75)
  correspond to various SUSY models.
(From Ref.~\protect\cite{AL-99}.)}
\label{alphabetacorr}
\end{figure}

% This is Figure 10
\begin{figure}
\vskip -2.0truein
\centerline{\epsfxsize 8.0 truein \epsfbox {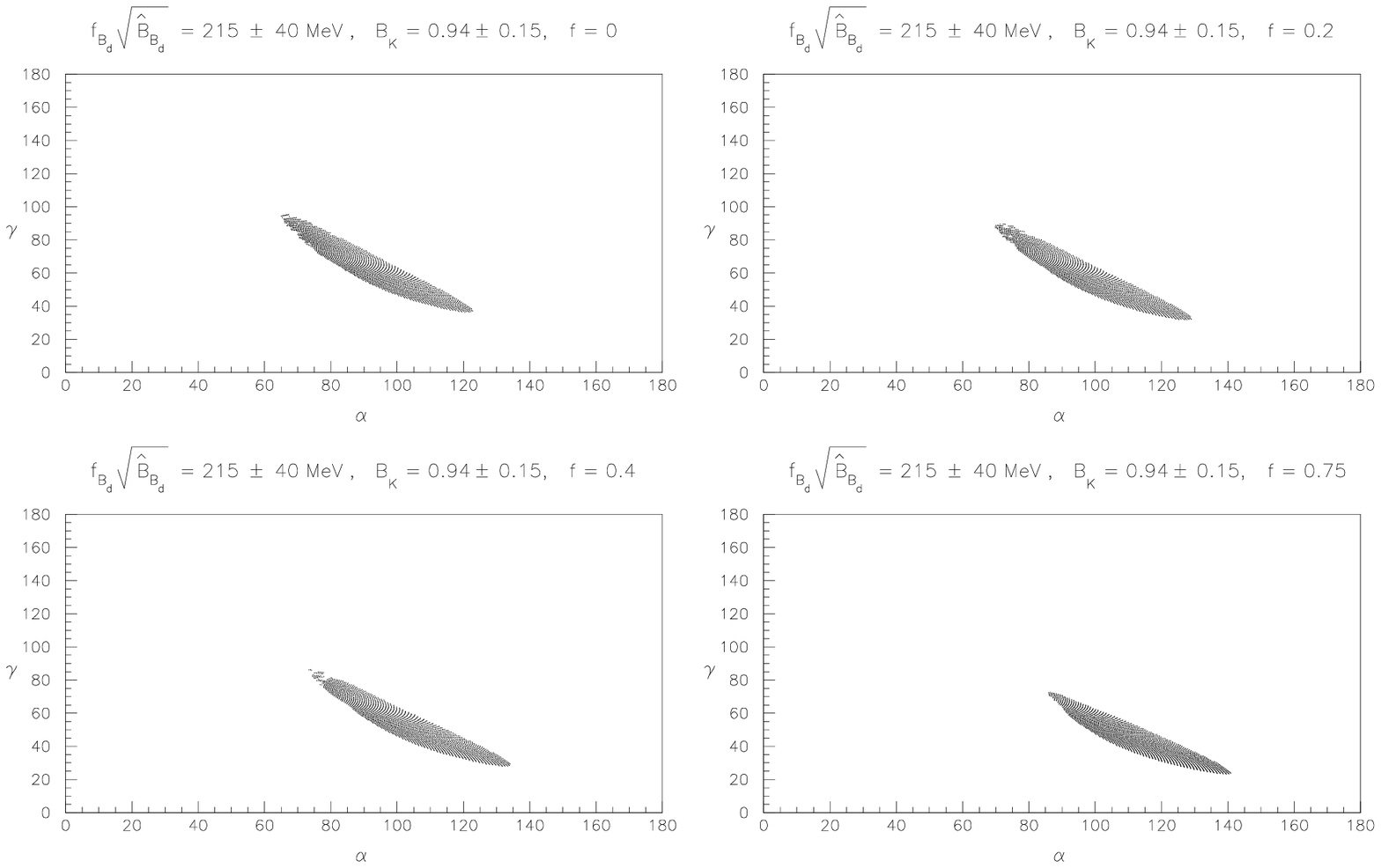}}
\vskip -4.7truein
\caption{\it Allowed 95\% C.L. region of the CP-violating quantities
  $\alpha$ and $\gamma$, from a fit to the data given in Table
  \protect{\ref{datatable}}. The upper left plot ($f=0$) corresponds
  to the SM, while the other plots ($f=0.2$, 0.4, 0.75) correspond to
  various SUSY models.
(From Ref.~\protect\cite{AL-99}.)}
\label{alphagammacorr}
\end{figure}
For the SUSY fits, we use the same program as for the SM fits, except
that the theoretical expressions for $\Delta M_d$, $\Delta M_s$ and
$\abseps$ are modified as in Eq.~(\ref{susyformel}). We compare the fits
for four representative values of the SUSY function $f$ --- 0, 0.2,
0.4 and 0.75 --- which are typical of the SM, minimal SUGRA models,
non-minimal SUGRA models, and non-SUGRA models with EDM constraints,
respectively.

% This is Figure 11
\begin{figure}
\vskip -1.0truein
\centerline{\epsfxsize 3.5 truein \epsfbox {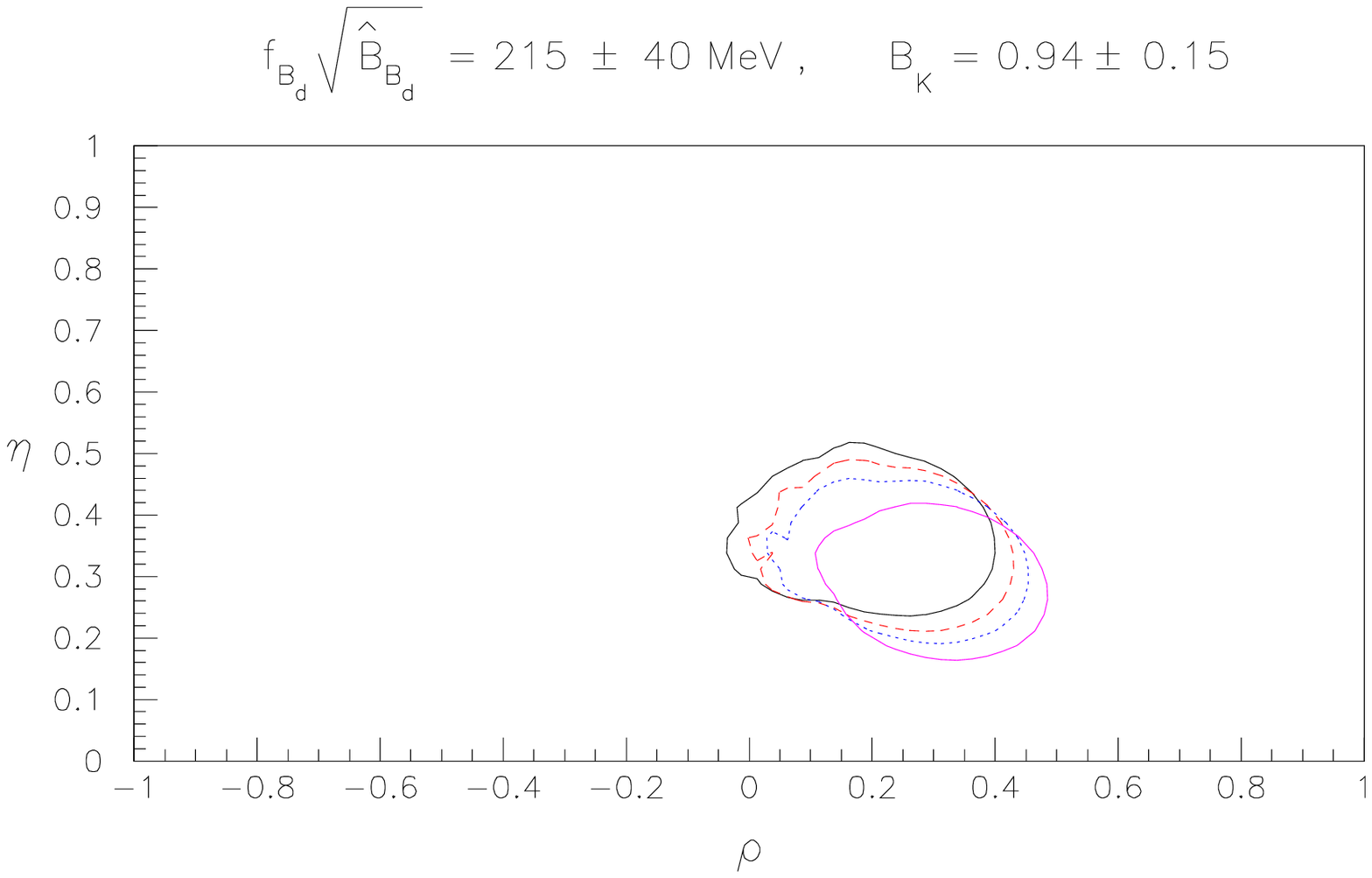}}
\vskip -1.4truein
\caption{\it Allowed 95\% C.L. region in $\rho$--$\eta$ space in the SM 
  and in SUSY models, from a fit to the data given in Table
  \protect{\ref{datatable}}. From left to right, the allowed regions
  correspond to $f=0$ (SM, solid line), $f=0.2$ (long dashed line),
  $f=0.4$ (short dashed line), $f=0.75$ (dotted line).
  (From Ref.~\protect\cite{AL-99}.)}
\label{sugratot}
\end{figure}

The allowed 95\% C.L. regions for these four values of $f$ are all
plotted in Fig.~\ref{sugratot}. As is clear from this figure, there is
still a considerable overlap between the $f=0$ (SM) and $f=0.75$
regions. However, there are also regions allowed for one value of $f$
which are excluded for another value. Thus a sufficiently precise
determination of the unitarity triangle might be able to exclude
certain values of $f$ (including the SM, $f=0$).

{}From Fig.~\ref{sugratot} it is clear that a measurement of the CP
angle $\beta$ will {\it not} distinguish among the various values of
$f$: even with the naked eye it is evident that the allowed range
for $\beta$ is roughly the same for all models. Rather, it is the
measurement of $\gamma$ or $\alpha$ which has the potential to rule
out certain values of $f$. As $f$ increases, the allowed region moves
slightly down and towards the right in the $\rho$--$\eta$ plane,
corresponding to smaller values of $\gamma$ (or equivalently, larger
values of $\alpha$). We illustrate this in Table~\ref{cpasym1}, where
we present the allowed ranges of $\alpha$, $\beta$ and $\gamma$, as
well as their central values (corresponding to the preferred values of
$\rho$ and $\eta$), for each of the four values of $f$. From this
Table, we see that the allowed range of $\beta$ is largely insensitive
to the model. Conversely, the allowed values of $\alpha$ and $\gamma$
do depend somewhat strongly on the chosen value of $f$. Note, however,
that one is not guaranteed to be able to distinguish among the various
models: as mentioned above, there is still significant overlap among
all four models. Thus, depending on what values of $\alpha$ and
$\gamma$ are obtained, we may or may not be able to rule out certain
values of $f$.

One point which is worth emphasizing is the correlation of $\gamma$
with $f$. This study clearly shows that large values of $f$ require
smaller values of $\gamma$. The reason that this is important is as
follows. The allowed range of $\gamma$ for a particular value of $f$
is obtained from a fit to all CKM data, even those measurements which
are unaffected by the presence of supersymmetry. Now, the size of
$\gamma$ indirectly affects the branching ratio for $B \to X_s
\gamma$: a larger value of $\gamma$ corresponds to a smaller value of
$|V_{ts}|$ through CKM unitarity. And this branching ratio is among
the experimental data used to bound SUSY parameters and calculate the
allowed range of $f$. Therefore, the above $\gamma$--$f$ correlation
indirectly affects the allowed values of $f$ in a particular SUSY
model, and thus must be taken into account in studies which examine
the range of $f$. For example, it is often the case that larger values
of $f$ are allowed for large values of $\gamma$. However, as we have seen
above, the CKM fits disfavour such values of $\gamma$.

\begin{table}
\hfil
\vbox{\offinterlineskip
\halign{&\vrule#&
 \strut\quad#\hfil\quad\cr
\noalign{\hrule}
height2pt&\omit&&\omit&&\omit&&\omit&&\omit&\cr 
& $f$ && $\alpha$ && $\beta$ && $\gamma$ && $(\alpha,\beta,\gamma)_{\rm cent}$ & \cr
height2pt&\omit&&\omit&&\omit&&\omit&&\omit&\cr 
\noalign{\hrule}
height2pt&\omit&&\omit&&\omit&&\omit&&\omit&\cr 
& $f=0$ (SM) && $65^\circ$ -- $123^\circ$ && $16^\circ$ -- $35^\circ$ && 
$36^\circ$ -- $97^\circ$ && $(93^\circ, 25^\circ, 62^\circ)$ & \cr 
& $f=0.2$ && $70^\circ$ -- $129^\circ$ && $16^\circ$ -- $34^\circ$ && 
$32^\circ$ -- $90^\circ$ && $(102^\circ, 24^\circ, 54^\circ)$ & \cr 
& $f=0.4$ && $75^\circ$ -- $134^\circ$ && $15^\circ$ -- $34^\circ$ && 
$28^\circ$ -- $85^\circ$ && $(110^\circ, 23^\circ, 47^\circ)$ & \cr 
& $f=0.75$ && $86^\circ$ -- $141^\circ$ && $13^\circ$ -- $33^\circ$ && 
$23^\circ$ -- $73^\circ$ && $(119^\circ, 22^\circ, 39^\circ)$ & \cr 
height2pt&\omit&&\omit&&\omit&&\omit&&\omit&\cr 
\noalign{\hrule}}}
\caption{\it Allowed 95\% C.L. ranges for the CP phases $\alpha$, $\beta$ 
and $\gamma$, as well as their central values, from the CKM fits in the
  SM $(f=0)$ and supersymmetric theories, characterized by the
  parameter $f$ defined in the text.}
\label{cpasym1}
\end{table}

For completeness, in Table~\ref{cpasym2} we present the corresponding
allowed ranges for the CP asymmetries $\sin 2\alpha$, $\sin 2\beta$
and $\sin^2 \gamma$. Again, we see that the allowed range of $\sin
2\beta$ is largely independent of the value of $f$. On the other hand,
as $f$ increases, the allowed values of $\sin 2\alpha$ become
increasingly negative, while those of $\sin^2 \gamma$ become smaller.

\begin{table}
\hfil
\vbox{\offinterlineskip
\halign{&\vrule#&
 \strut\quad#\hfil\quad\cr
\noalign{\hrule}
height2pt&\omit&&\omit&&\omit&&\omit&\cr
& $f$ && $\sin 2\alpha$ &&
$\sin 2\beta$ && $\sin^2 \gamma$ & \cr
height2pt&\omit&&\omit&&\omit&&\omit&\cr
\noalign{\hrule}
height2pt&\omit&&\omit&&\omit&&\omit&\cr
& $f=0$ (SM) && $-$0.91 -- 0.77 && 0.53 -- 0.94 && 0.35 -- 1.00 & \cr
& $f=0.2$ && $-$0.98 -- 0.65 && 0.52 -- 0.93 && 0.28 -- 1.00 & \cr
& $f=0.4$ && $-$1.00 -- 0.50 && 0.49 -- 0.93 && 0.22 -- 0.99 & \cr
& $f=0.75$ && $-$1.00 -- 0.14 && 0.45 -- 0.91 && 0.16 -- 0.91 & \cr
height2pt&\omit&&\omit&&\omit&&\omit&\cr
\noalign{\hrule}}}
\caption{\it Allowed 95\% C.L. ranges for the CP asymmetries $\sin 2\alpha$, 
$\sin 2\beta$ and $\sin^2 \gamma$, from the CKM fits in the SM $(f=0)$ and
  supersymmetric theories, characterized by the parameter $f$ defined
  in the text.}
\label{cpasym2}
\end{table}

The allowed (correlated) values of the CP angles for various values of
$f$ can be clearly seen in Figs.~\ref{alphabetacorr} and
\ref{alphagammacorr}. As $f$ increases from 0 (SM) to 0.75, the change
in the allowed $\sin 2\alpha$--$\sin 2\beta$ (Fig.~\ref{alphabetacorr})
and $\alpha$--$\gamma$ (Fig.~\ref{alphagammacorr}) regions is quite
significant.

In Sec.~2.1, we noted that $|V_{ub}/V_{cb}|$, ${\hat B}_K$ and
$\fbd\sqrt{\hat{B}_{B_d}}$ are very important in defining the allowed
region in the $\rho$--$\eta$ plane. At present, these three quantities
have large errors, which are mostly theoretical in nature. Let us
suppose that our theoretical understanding of these quantities
improves, so that the errors are reduced by a factor of two, i.e.
\begin{eqnarray}
\left\vert { V_{ub} \over V_{cb} } \right\vert & = & 0.093 \pm 0.007 ~, \nn\cr
\hat{B}_K & = & 0.94 \pm 0.07 ~, \nn\cr
\fbd\sqrt{\hat{B}_{B_d}} & = & 215 \pm 20~{\rm MeV} ~.
\label{newdata}
\end{eqnarray}
How would such an improvement affect the SUSY fits?

\begin{figure}
\vskip -1.0truein
\centerline{\epsfxsize 3.5 truein \epsfbox {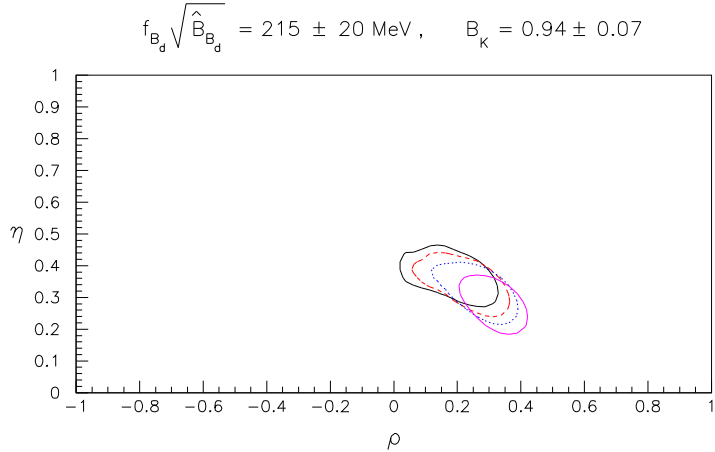}}
\vskip -1.4truein
\caption{\it Allowed 95\% C.L. region in $\rho$--$\eta$ space in the SM 
  and in SUSY models, from a fit to the data given in Table
  \protect{\ref{datatable}}, with the (hypothetical) modifications
  given in Eq.~(\protect\ref{newdata}). From left to right, the
  allowed regions correspond to $f=0$ (SM, solid line), $f=0.2$ (long
  dashed line), $f=0.4$ (short dashed line), $f=0.75$ (dotted line).
  (From Ref.~\protect\cite{AL-99}.)}
\label{sugratot_fut}
\end{figure}

We present the allowed 95\% C.L. regions ($f=0$, 0.2, 0.4, 0.75) for
this hypothetical situation in Fig.~\ref{sugratot_fut}. Not
surprisingly, the regions are quite a bit smaller than in
Fig.~\ref{sugratot}. More importantly for our purposes, the regions
for the different values of $f$ have become more separated from one
another. That is, although there is still a region where all four $f$
values are allowed, precise measurements of the CP angles have a
better chance of ruling out certain values of $f$.

% This is Figure 12

In Table~\ref{cpasym1_fut} we present the allowed ranges of $\alpha$,
$\beta$ and $\gamma$, as well as their central values, for this
scenario. Table~\ref{cpasym2_fut} contains the corresponding allowed
ranges for the CP asymmetries $\sin 2\alpha$, $\sin 2\beta$ and
$\sin^2 \gamma$. The allowed $\sin 2\alpha$--$\sin 2\beta$ and
$\alpha$--$\gamma$ correlations can be seen in \cite{AL-99}.
As is consistent with the smaller regions of
Fig.~\ref{sugratot}, the allowed (correlated) regions are
considerably reduced compared to Figs.~\ref{alphabetacorr} and
\ref{alphagammacorr}. As before, although the measurement of $\beta$
will not distinguish among the various values of $f$, the measurement
of $\alpha$ or $\gamma$ may.

Indeed, the assumed reduction of errors in Eq.~(\ref{newdata})
increases the likelihood of this happening. For example, consider
again Table~\ref{cpasym1}, which uses the original data set of
Table~\ref{datatable}. Here we see that $65^\circ \le \alpha \le
123^\circ$ for $f=0$ and $86^\circ \le \alpha \le 141^\circ$ for
$f=0.75$. Thus, if experiment finds $\alpha$ in the range
$86^\circ$--$123^\circ$, one cannot distinguish the SM ($f=0$) from
the SUSY model with $f=0.75$. However, consider now
Table~\ref{cpasym1_fut}, obtained using data with reduced errors.
Here, $67^\circ \le \alpha \le 116^\circ$ for $f=0$ and $97^\circ \le
\alpha \le 137^\circ$ for $f=0.75$. Now, it is only if experiment
finds $\alpha$ in the range $97^\circ$--$116^\circ$ that one cannot
distinguish $f=0$ from $f=0.75$. But this range is quite a bit smaller
than that obtained using the original data. This shows how an
improvement in the precision of the data can help not only in
establishing the presence of new physics, but also in distinguishing
among various models of new physics.

\begin{table}
\hfil
\vbox{\offinterlineskip
\halign{&\vrule#&
 \strut\quad#\hfil\quad\cr
\noalign{\hrule}
height2pt&\omit&&\omit&&\omit&&\omit&&\omit&\cr 
& $f$ && $\alpha$ && $\beta$ && $\gamma$ && $(\alpha,\beta,\gamma)_{\rm cent}$ & \cr
height2pt&\omit&&\omit&&\omit&&\omit&&\omit&\cr 
\noalign{\hrule}
height2pt&\omit&&\omit&&\omit&&\omit&&\omit&\cr 
& $f=0$ (SM) && $67^\circ$ -- $116^\circ$ && $20^\circ$ -- $30^\circ$ && 
$42^\circ$ -- $90^\circ$ && $(93^\circ, 24^\circ, 63^\circ)$ & \cr 
& $f=0.2$ && $74^\circ$ -- $124^\circ$ && $19^\circ$ -- $29^\circ$ && 
$36^\circ$ -- $82^\circ$ && $(102^\circ, 24^\circ, 54^\circ)$ & \cr 
& $f=0.4$ && $83^\circ$ -- $130^\circ$ && $18^\circ$ -- $29^\circ$ && 
$31^\circ$ -- $73^\circ$ && $(110^\circ, 23^\circ, 47^\circ)$ & \cr 
& $f=0.75$ && $97^\circ$ -- $137^\circ$ && $16^\circ$ -- $28^\circ$ && 
$26^\circ$ -- $59^\circ$ && $(119^\circ, 22^\circ, 39^\circ)$ & \cr 
height2pt&\omit&&\omit&&\omit&&\omit&&\omit&\cr 
\noalign{\hrule}}}
\caption{\it Allowed 95\% C.L. ranges for the CP phases $\alpha$, $\beta$ 
and $\gamma$, as well as their central values, from the CKM fits in the
  SM $(f=0)$ and supersymmetric theories, characterized by the
  parameter $f$ defined in the text. We use the data given in Table
  \protect{\ref{datatable}}, with the (hypothetical) modifications
  given in Eq.~(\protect\ref{newdata}).}
\label{cpasym1_fut}
\end{table}

\begin{table}
\hfil
\vbox{\offinterlineskip
\halign{&\vrule#&
 \strut\quad#\hfil\quad\cr
\noalign{\hrule}
height2pt&\omit&&\omit&&\omit&&\omit&\cr
& $f$ && $\sin 2\alpha$ &&
$\sin 2\beta$ && $\sin^2 \gamma$ & \cr
height2pt&\omit&&\omit&&\omit&&\omit&\cr
\noalign{\hrule}
height2pt&\omit&&\omit&&\omit&&\omit&\cr
& $f=0$ (SM) && $-$0.80 -- 0.71 && 0.64 -- 0.86 && 0.44 -- 1.00 & \cr
& $f=0.2$ && $-$0.93 -- 0.53 && 0.61 -- 0.85 && 0.34 -- 0.98 & \cr
& $f=0.4$ && $-$0.99 -- 0.23 && 0.57 -- 0.85 && 0.27 -- 0.91 & \cr
& $f=0.75$ && $-$1.00 -- $-$0.23 && 0.52 -- 0.83 && 0.19 -- 0.73 & \cr
height2pt&\omit&&\omit&&\omit&&\omit&\cr
\noalign{\hrule}}}
\caption{\it Allowed 95\% C.L. ranges for the CP asymmetries $\sin 2\alpha$, 
$\sin 2\beta$ and $\sin^2 \gamma$, from the CKM fits in the SM $(f=0)$ and
  supersymmetric theories, characterized by the parameter $f$ defined
  in the text. We use the data given in Table \protect{\ref{datatable}}, 
  with the (hypothetical) modifications given in 
Eq.~(\protect\ref{newdata}).} \label{cpasym2_fut}
\end{table}

\section{Conclusions}

In the very near future, CP-violating asymmetries in $B$ decays will
be measured at $B$-factories, HERA-B and hadron colliders. Such measurements
will give us crucial information about the interior angles $\alpha$,
$\beta$ and $\gamma$ of the unitarity triangle. If we are lucky, there
will be an inconsistency in the independent measurements of the sides
and angles of this triangle, thereby revealing the presence of new
physics.

If present, this new physics will affect $B$ decays principally
through new contributions to $B^0$--${\overline{B^0}}$ mixing. If these
contributions come with new phases (relative to the SM), then the CP
asymmetries can be enormously shifted from their SM values. In this
case there can be huge discrepancies between measurements of the
angles and the sides, so that the new physics will be easy to find.

A more interesting possibility, from the point of view of making
predictions, are models which contribute to $B^0$--${\overline{B^0}}$
mixings and $\abseps$, but without new phases. One type of new physics
which does just this is supersymmetry (SUSY). There are some SUSY
models which do contain new phases, but they suffer from the problem
described above: lack of predictivity. However, there is also a large
class of SUSY models with no new phases. In this paper we have
concentrated on these models.

In these models, there are new, supersymmetric contributions to
\kkbar, \bdbdbar\ and \bsbsbar\ mixing. The key ingredient in our
analysis is the fact that these contributions, which add
constructively to the SM, depend on the SUSY parameters in essentially
the same way. That is, so far as an analysis of the unitarity triangle
is concerned, there is a single parameter, $f$, which characterizes
the various SUSY models within this class of models ($f=0$ corresponds
to the SM). For example, the values $f=0.2$, 0.4 and 0.75 are found in
minimal SUGRA models, non-minimal SUGRA models, and non-SUGRA models
with EDM constraints, respectively.

We have therefore updated the profile of the unitarity triangle in
both the SM and some variants of the MSSM. We have used the latest 
experimental 
data on $|V_{cb}|$, $|V_{ub}/V_{cb}|$, $\Delta M_d$ and $\Delta M_s$, as
well as the latest theoretical estimates (including errors) of
$\hat{B}_K$, $\fbd\sqrt{\hat{B}_{B_d}}$ and $\xi_s \equiv
\fbd\sqrt{\hat{B}_{B_d}}/\fbs\sqrt{\hat{B}_{B_s}}$. In addition to
$f=0$ (SM), we considered the three SUSY values of $f$: 0.2, 0.4 and
0.75.

We first considered the profile of the unitarity triangle in the SM,
shown in Fig.~\ref{rhoeta1}. At present, the allowed ranges for the CP
angles at 95\% C.L. are
\beq
65^\circ \le \alpha \le 123^\circ ~,~~
16^\circ \le \beta \le 35^\circ ~,~~
36^\circ \le \gamma \le 97^\circ ~,
\eeq
or equivalently,
\beq
-0.91 \le  \sin 2\alpha  \le 0.77 ~,~
0.52  \le  \sin 2\beta  \le 0.94  ~,~
0.35  \le  \sin^2 \gamma  \le 1.00 ~.
\eeq
We have also performed CKM fits for the superweak model. This is done by
leaving out the constraint from $\abseps$. The resulting allowed unitarity
triangle now depends on the value of $\fbd\sqrt{\hat{B}_{B_d}}$. With the
present estimate $\fbd\sqrt{\hat{B}_{B_d}}=215 \pm 40$ MeV, the superweak
case, i.e. $\eta=0$, is ruled out at 95\% C.L. However, unless the 
theoretical error on this quantity is reduced, the resulting value of
$\eta$ has a large uncertainty.  

We then compared the SM with the different SUSY models. The result can
be seen in Fig.~\ref{sugratot}. As $f$ increases, the allowed region
moves slightly down and to the right in the $\rho$--$\eta$ plane. The
main conclusion from this analysis is that the measurement of the CP
angle $\beta$ will not distinguish among the SM and the various SUSY
models -- the allowed region of $\beta$ is virtually the same in all
these models. On the other hand, the allowed ranges of $\alpha$ and
$\gamma$ do depend on the choice of $f$. For example, larger values of
$f$ tend to favour smaller values of $\gamma$. Thus, with measurements
of $\gamma$ or $\alpha$, we may be able to rule out certain values of
$f$ (including the SM, $f=0$). However, we also note that there is no
guarantee of this happening -- at present there is still a significant
region of overlap among all four models.

Finally, we also considered a hypothetical future data set in which
the errors on $|V_{ub}/V_{cb}|$, $\hat{B}_K$ and
$\fbd\sqrt{\hat{B}_{B_d}}$, which are mainly theoretical, are reduced
by a factor of two. For two of these quantities ($|V_{ub}/V_{cb}|$ and
$\fbd\sqrt{\hat{B}_{B_d}}$), this has the effect of reducing the
uncertainty on the sides of the unitarity triangle by the same factor.
The comparison of the SM and SUSY models is shown in
Fig.~\ref{sugratot_fut}. As expected, the allowed regions for all
models are quite a bit smaller than before. Furthermore, the regions
for different values of $f$ have become more separated, so that
precise measurements of the CP angles have a better chance of ruling
out certain values of $f$.

\section*{Acknowledgements}
We thank Laksana Tri Handoko for his help in producing Figs.~7 and 8. 

%%%%%%%%%%%%%%%%%%%%% REFERENCES %%%%%%%%%%%%%%%%%%%%%%%%%%%%%%%%

\end{document}